\documentclass[11pt,journal,draftcls,letterpaper,onecolumn]{IEEEtran}

 \usepackage[utf8]{inputenc}
 \usepackage{amsmath}
 \usepackage{amsfonts}
 \usepackage{amssymb}
 \usepackage{graphicx}
 \usepackage{color}
 \usepackage{blindtext}
 \usepackage{scalefnt}
 \usepackage{bbm}
 \usepackage{algorithm}
\usepackage{algorithmic}

 \def\vrho{\overline{\rho}}
 
 \def\n{\mathtt{n}}

 \def\y{\boldsymbol{y}}
 \def\fraction{\nu}
 \def\ep{\epsilon}
  
 \def\rate{\mathtt{R}}
 \def\length{\mathtt{n}}
 \def\edges{\mathtt{E}}
 \def\cnodes{\mathtt{c}}
 \def\vlambda{\overline{\lambda}}
 
 \def\vrho{\overline{\rho}}
 \def\dist{\mathtt{d}}
 \def\rows{\mathtt{k}}
 \def\rowsv{\mathtt{m}}
 \def\regC{\mathcal{C}_{J,K,\nu}}

 \def\rowavg{\rows_{\text{avg}}}
 
 \def\rateC{\rate(\fraction)}
 \def\rateDC{\rate(\alpha,\beta)}

\def\t{^{(\tau)}}
 \def\l{^{(\ell)}}

\usepackage{tikz}
\usetikzlibrary{fit,positioning}
\usetikzlibrary{shapes,matrix,decorations.markings,arrows}
\usetikzlibrary{graphs}
\usepackage{pgfplots}
\usepackage{pgfplotstable}
\usepackage{filecontents}
\usetikzlibrary{plotmarks}
\usetikzlibrary{positioning}
\usepgfplotslibrary{fillbetween}

\usetikzlibrary{external} 
 \def\th{\ep^*}
  
    \def\PDth{\epsilon_0}
 
 \newtheorem{theorem}{Theorem}
\newtheorem{lemma}[theorem]{Lemma}

\newcommand{\nV}[1]{\textcolor{black}{#1}}

\title{A Probabilistic Peeling Decoder to Efficiently Analyze Generalized LDPC Codes Over the BEC}

\author{
\IEEEauthorblockN{Yanfang Liu, Pablo M. Olmos, Tobias Koch\\}
\IEEEauthorblockA{
	Universidad Carlos III de Madrid \& Gregorio Mara\~n\'on Health Research Institute\\
	Email: \texttt{\{vivian,olmos,koch\}@tsc.uc3m.es}
	}
	
	\thanks{
}
}

\usepackage{pdfpages}

\begin{document}

\maketitle

\begin{abstract}

In this paper, we analyze the tradeoff between coding rate and asymptotic performance of a class of generalized low-density parity-check (GLDPC) codes constructed by including a certain fraction of generalized constraint (GC) nodes in the graph. The rate of the GLDPC ensemble is bounded using classical results on linear block codes, namely Hamming bound and Varshamov bound.   We also study the impact of the decoding method used at GC nodes. To incorporate both bounded-distance (BD) and Maximum Likelihood (ML) decoding at GC nodes into our analysis without resorting on multi-edge type of degree distributions (DDs), we propose the probabilistic peeling decoding (P-PD) algorithm, which models the decoding step at every GC node as an instance of a Bernoulli random variable with a successful decoding probability that depends on both the GC block code as well as its decoding algorithm. The P-PD asymptotic performance over the BEC can be efficiently predicted using standard techniques for LDPC codes such as density evolution (DE) or the differential equation method. Furthermore, for a class of GLDPC ensembles, we demonstrate  that the simulated P-PD performance accurately predicts the actual performance of the GLPDC code under ML decoding at GC nodes. We illustrate our analysis for GLDPC code ensembles with regular and irregular DDs. In all cases, we show that a large fraction of GC nodes is required to reduce the original gap to capacity, but the optimal fraction is strictly smaller than one. We then consider techniques to further reduce the gap to capacity by means of random puncturing, and the inclusion of a certain fraction of generalized variable nodes in the graph.


\end{abstract}

\section{Introduction}
{\let\thefootnote\relax\footnotetext{
This work has been funded in part by the Spanish Ministerio de Econom\'ia y Competitividad and the Agencia Espa\~nola de Investigaci\'on under Grant TEC2016-78434-C3-3-R (AEI/FEDER, EU) and by the Comunidad de Madrid in Spain under Grant S2103/ICE-2845. T. Koch has further received funding from the European Research Council (ERC) under the European Union's Horizon 2020 research and innovation programme (grant agreement number 714161), from the 7th European Union Framework Programme under Grant 333680, and from the Spanish Ministerio de Econom\'ia y Competitividad under Grants TEC2013-41718-R and RYC-2014-16332. Pablo M. Olmos has further received funding from the Spanish Ministerio de Econom\'ia y Competitividad under Grant IJCI-2014-19150. This paper was presented in part at 2017 IEEE International Symposium on Information Theory.}}

Generalized low-density parity-check (GLDPC)  block codes were first proposed by Tanner  \cite{Tanner81}.  In contrast to standard LDPC codes, which are represented by bipartite Tanner graphs where variable nodes and single parity-check (SPC) nodes are connected according to a given degree distribution (DD), in GLDPC codes the SPC nodes in the graph are replaced by generalized constraint (GC) nodes \cite{Tanner81}. The sub-code associated to each GC node is referred to as the component code. Examples of component codes used in the GLDPC literature are Hamming  codes \cite{Lentmaier99}, Hadamard codes \cite{Yue07} or expurgated random codes \cite{Liva08,Paolini10}. For powerful component codes, GLDPC codes have many potential advantages, including improved performance in noisy channels, fast convergence speed \cite{Mulholland15} and low error floor \cite{Liva08,Mitchell13GLDPC}.

Upon selecting a particular class of component codes, the DD of the GLDPC code ensemble can be optimized, and near-capacity iterative decoding thresholds can be achieved \cite{Lentmaier99,Liva08,adr11}. Capacity-achieving GLDPC code ensembles can also  be obtained by spatially-coupling GLDPC block codes with regular DDs \cite{Lentmaier10GLDPC,Jian2012}. Furthermore, the asymptotic exponents of the weight/stopping set spectrum for irregular and spatially-coupled GLDPC ensembles have been derived in \cite{Mitchell13GLDPC} and \cite{Paolini13}, respectively. Based on these works, it is possible to design asymptotically good GLDPC code ensembles to achieve capacity-approaching iterative decoding thresholds and a minimum distance that grows linearly with the blocklength.

In this paper, we analyze GLDPC code ensembles using a different approach. Instead of selecting a particular class of component codes and optimizing the graph DD, we are  interested in analyzing the tradeoff between coding rate and iterative decoding threshold of GLDPC code ensembles with fixed DD, referred to as the base DD, as we increase the fraction $\fraction$ of GC nodes in the graph.  This approach is novel in the literature and we believe it is appealing from a design perspective, since one might be interested in introducing a certain amount of GC nodes in the Tanner graph of a given LDPC code, aiming at reducing the gap to channel capacity at the resulting coding rate, and at the same time  improving \color{black} the minimum distance of the code and thus the error floor. 


For the BEC, iterative decoding of graph-based codes, such as LDPC or GLDPC codes, can be performed by means of \emph{peeling decoding} (PD) algorithms \cite{Luby01,Urbanke08,Olmos15}, which iteratively remove  from the Tanner graph variable nodes whose value is known. As a result, the decoding process yields a sequence of graphs whose mean coincides with the asymptotic (in the blocklength) evolution of the ensemble. Furthermore, this evolution can be computed by solving a particular set of differential equations \cite{Luby01}. In the case of GLDPC codes the derivation of such differential equations requires to specify in advance the DD of the graph, and a description of what  kind of  erasure patterns are locally decodable at any GC node, which depends on both the component codes and the corresponding decoding algorithm. In fact, the resulting decoding threshold of GLDPC codes heavily depends on this latter point \cite{Yue07,Paolini10,Olmos15}. For instance, as we demonstrated in this paper, for a $(2,7)$ base DD in which all check nodes are $(7,4)$-Hamming GC nodes, the asymptotic threshold over the BEC is $\epsilon^*\approx 0.7025$ if maximum likelihood (ML) decoding is performed at each GC node. However, it drops to $\epsilon^*\approx 0.5135$ if suboptimal bounded distance (BD) decoding  is used instead of ML. In both cases, the coding rate is exactly the same. The reason for this difference in performance is that BD-decoded GC nodes only resolve erasure patterns up to degree $\dist-1$, where $\dist$ is the minimum distance of the component code, whereas ML-decoded GC nodes can resolve a subset of erasure patterns of degree above $\dist-1$. Note, however, that this improvement of performance comes at the cost of higher complexity. Let $K$ denote the blocklength of the component code. For the BEC, the ML-decoding complexity at GC nodes is of order $\mathcal{O}(K^3)$, since it is equivalent to solving a system of binary linear equations \cite{Burshtein04}. 

While deriving the asymptotic differential equations to analyze PD with BD decoding at GC nodes (BD-PD for short) follows a straightforward extension of the standard PD differential equations for LDPC codes \cite{Luby01}, the GLDPC asymptotic analysis of PD under ML-decoded component codes (ML-PD, for short) requires the use of multi-edge-type DDs \cite{Urbanke08-2} to track down all possible decodable erasure patterns at GC nodes \cite{Lentmaier10GLDPC, Olmos15}. As a consequence, the list of code parameters to jointly optimize becomes cumbersome. Specifically, the parameters include the description of the multi-edge DD, the position of GC nodes in the graph, the edge labelling  at every GC node used to determine positions in the component block code, and the list of locally ML-decodable erasure patterns.  In \cite{Paolini10}, the authors were able to incorporate ML-decoded GC nodes without resorting to multi-edge type DDs by analyzing the  GLDPC average performance using extrinsic information (EXIT) charts when each GC node in the graph is selected at random within the family of block component codes with fixed block length and minimum distance larger than 2. This approach has a design caveat though, as it does neither allow the use of a single type of component codes, nor to narrow down the family of component codes by fixing the minimum distance.

In this paper, we propose an analysis methodology that allows to easily incorporate into the PD algorithm ML-decoded GC nodes with specific properties, such a particular value of the minimum distance $\dist$ or how many erasure patterns beyond minimum distance it can decode. We develop a probabilistic description of all components of the GLDPC code, namely the base DD, the presence of GC nodes in the graph, and the decoding method implemented at GC nodes. Regarding the latter aspect, we parameterize the decoding capabilities of at every node with a blocklength-$K$ component code by a vector $(p_1,p_2,...,p _K)$, where $p_{w} \in [0,1]$, $w\in\{1, \ldots,K\}$, is the probability that a weight-$w$ erasure pattern chosen at random is decodable. Thus, $p_{w}$ is the fraction of decodable weight-$w$ erasure patterns. Note that if we take $p_{w} =1$ for $w\leq \dist-1$ and $p_{w} = 0$ for $w = \{\dist,\ldots,K\}$, we recover BD-PD. We show how to properly incorporate such a probabilistic description of component codes into the PD algorithm, and denote the resulting algorithm as \emph{probabilistic PD} (P-PD). Due to its probabilistic nature, the asymptotic analysis of P-PD does not require the use of multi-edge type DDs. We show by computer simulations that the P-PD performance accurately predicts the actual GLDPC performance when ML decoding is performed at GC nodes. We note that the proposed techniques are valid  for binary GLDPC codes and that we do not consider non-binary LDPC codes \cite{MacKay03}, which can also be considered a special class of GLDPC codes. 

The performance predicted using P-PD is valid for any linear component code of blocklength-$K$ and decoding profile $(p_1,p_2,...,p _K)$. To analyze a family of linear component codes of blocklength-$K$ and minimum distance $\dist$, we employ two bounds to compute the GLDPC coding rate. The Hamming or sphere-packing bound  \cite{MacWilliams77} is used to determine a converse bound on the rate of the GLDPC code ensemble as a function of a triplet of $(\fraction,\dist, K)$. The Varshamov bound is  considered to determine an achievable rate of the GLDPC code ensemble \cite{Huffman03}. In many scenarios of interest, we show that these bounds are sufficiently tight and thus relevant for the code designer. 

By employing a probabilistic description of the decoding capabilities at GC nodes, we are able to analyze a large class of GLDPC code ensembles and beyond-BD decoding methods with a fairly small set of parameters. We illustrate our analysis for both regular GLDPC code ensembles using $(2,6)$, $(2,7)$, $(2,8)$ and $(2,15)$ base DDs and irregular GLDPC code ensembles with similar graph densities \cite{Paolini08, Guan17}. To obtain realistic values for the coding capabilities of the component codes, we have performed an exhaustive search of linear block codes of lengths $r \in [6, 7, 8, 15]$, including Hamming codes, Cyclic codes, Quasi Cyclic codes and Cordaro-Wagner Codes, and tabulated their corresponding description in terms of minimum distance $\dist$ and $(p_1,p_2,...,p _K)$. In all cases, we show that a large fraction of GC nodes is required in the GLDPC graph to  reduce the original gap to capacity. However, the closest gap to capacity is not achieved at $\fraction=1$, but a smaller value must be used. Namely, there exists a critical $\fraction^*$ value for which the gap to capacity is minimum. Furthermore, the best results are obtained for high-rate component codes, suggesting that the use of very powerful component codes does not pay off, since the gain in threshold does not compensate for the severe decrease of the GLDPC code rate. Furthermore, we include into our analysis the weight spectral analysis of GLDPC ensembles in \cite{Paolini13} to explore the range of $\fraction$ values for which the GLDPC ensembles reduce the original gap to capacity and at the same time maintain a linear growth of the minimum distance with the block length. 
%

Finally, we illustrate how to incorporate further design techniques that can help to reduce the gap to capacity of the code ensembles. Specifically, we discuss both random puncturing \cite{Mitchell15} and a simple class of doubly generalized LDPC (DG-LDPC) codes \cite{Wang06,Yige09}.  
In general, the methodology presented in this paper is flexible and decouples the problems of bounding the GLDPC coding rate and the asymptotic analysis of the ensemble. In this regard, broader classes of component codes at variable nodes and GC nodes could also be incorporated in a systematic way. 

The paper is organized as follows. In Section \ref{codes}, we introduce GLDPC code ensembles and the notation used to characterize the DDs. Sections \ref{Peeling} and \ref{section_four} present the decoding algorithm and its asymptotic analysis. In Section \ref{SecV} we bound the GLDPC code rate and analyze the rate-threshold tradeoff as a function of the fraction $\fraction$ of GC nodes in the graph. The behavior of the GLDPC code ensembles with specific component codes is analyzed in Section \ref{MLPD}. Finally, Sections  \ref{rateadaptation} and  \ref{doubly} consider further techniques to improve the asymptotic behavior of the code ensemble, by means of random puncturing and generalized variable nodes. We conclude the paper in Section \ref{end} with a discussion of our results.
 
\section{GLDPC ensembles} \label{codes}
In this section, we introduce the GLDPC code ensembles that will be analyzed in the rest of the paper and the notation used to define their DD. 
\subsection{Degree distribution}
\begin{figure}[t]
\centering
\includegraphics{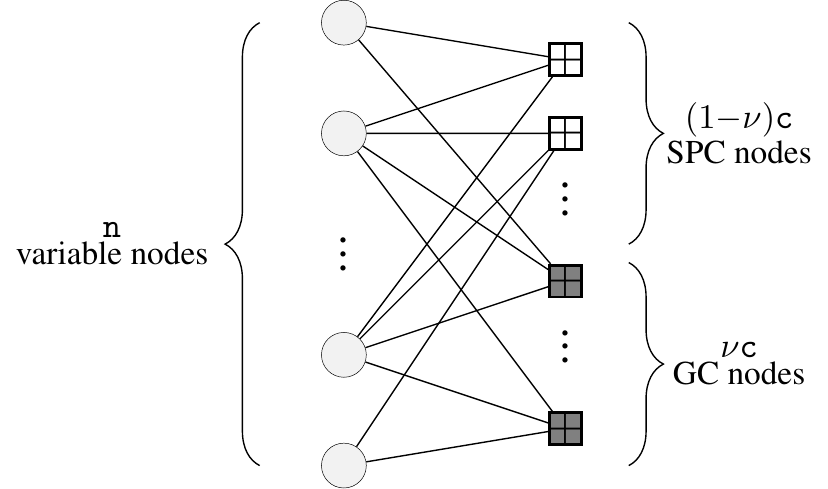}
\caption{Tanner graph of a GLDPC code.}\label{GLDPCgraph}
\end{figure}

As illustrated in Fig. \ref{GLDPCgraph}, the Tanner graph of every member in the ensemble contains $\length$ variable nodes (coded bits) and $\cnodes$ parity-check nodes, among which a fraction $\fraction$ corresponds to GC nodes while the rest corresponds to SPC nodes. We denote by $\edges$ the  number of edges in the Tanner graph and we define the degree of a node  as the number of edges connected to it. 

The DD of the ensemble is characterized as follows. The vector $\vlambda = (\lambda_1,\lambda_2,...,\lambda _J)$ is the \emph{left}  DD, where $\lambda_i$ represents the  fraction of edges (w.r.t. $\edges$) connected to a variable node of degree $i$. Given $\vlambda$, $\n$ and $\edges$ are related by \cite{Urbanke08-2}
\begin{align}
\n=\edges\sum_{i=1}^{J}\lambda_i/i.
\end{align}

The \emph{right} DD is defined by two vectors $\vrho_p= (\rho_{p1},\rho_{p2},...,\rho _{pK})$ and $\vrho_c = (\rho_{c1},\rho_{c2},..., \rho _{cK})$, where $\rho_{pj}$ denotes the  fraction of edges (w.r.t. $\edges$) connected to a  SPC node that has degree $j$ and $\rho_{cj}$ denotes the fraction of edges (w.r.t. $\edges$) connected to a GC node that has degree $j$. Throughout the paper, we use the subscript $p$ for any DD component related to standard \emph{parity} check nodes and the subscript $c$ for any DD component related to generalized \emph{component} codes. The DD is then characterized by the tuple $(\vlambda,\vrho_p,\vrho_c,\fraction)$ and the ensemble of codes generated by this DD is denoted by $\mathcal{C}_{\vlambda,\vrho_p,\vrho_c,\fraction}$.
Since the fraction of GC nodes in the graph is $\fraction$,  the following must hold:
\begin{align}
\fraction=\frac{\sum_{j=1}^{K}\rho_{cj}/j}{\sum_{u=1}^{K}(\rho_{cu}+\rho_{pu})/u}.
\end{align}

For simplicity, we restrict the most of our analysis to the class of GLDPC ensembles characterized by variable nodes with constant degree $J$ and SPC and GC nodes with constant degree and $K$.  The Tanner graph of any code in this ensemble contains  $\n$ variable nodes, $\edges=J\n$ edges, $\fraction\frac{J}{K}\n$ GC nodes, and $(1-\fraction)\frac{J}{K}\n$ SPC nodes. The DD of the GLDPC codes is characterized by the triple $(J,K,\fraction)$,  and the ensemble of codes generated by this DD is denoted by $\regC$. The DD of the LDPC ensemble obtained by taking $\fraction=0$ is defined as the \emph{base DD}, and the corresponding LDPC  code ensemble is referred to as the \emph{base ensemble}. The coding rate of the base ensemble is denoted by $\rate_0$ and can be computed as:
\begin{align}\label{R0}
\rate_0=1-\frac{J}{K}.
\end{align}

Finally, we assume that the incoming edges to every degree-$K$ GC node are assigned uniformly at random to each position of the component code.

\subsection{The coding rate of the $\regC$ ensemble}
\label{rate}

As discussed in the introduction of the paper, we propose tools to analyze the decoding performance of GLDPC under ML-decoded GC nodes that do not require to set in advance a specific component code to be used as the GC nodes. Instead, we consider the family of linear block codes with blocklength $K$ and minimum distance $\dist$, and we use the  classical results on linear block codes to bound the coding rate of the GLDPC code ensembles.  


Let $\rows^{(\ell)}\in\mathbb{N}^+$, $\ell=1,\ldots, \fraction\edges/K$, be the number of rows in the parity-check matrix associated with the component code of the $\ell$-th GC node. 
\lemma{
The design rate $\rateC$ of the $\regC$ ensemble is
 \begin{align}\label{rate2}
\rateC=\rate_0-\fraction(1-\rate_0)(\rowavg-1),
\end{align}
where $\rowavg \triangleq (\nu\frac{\edges}{K})^{-1}\sum_{\ell=1}^{\nu\frac{\edges}{K}}\rows^{(\ell)}$ is the average number of rows in the parity-check matrix of the component codes.
}
\begin{IEEEproof}
Any SPC node in the Tanner graph accounts for a single row in the parity-check matrix of the GLDPC code, and any GC node accounts for $\rows^{(\ell)}$ rows. Thus, the design rate $\rateC$ is given by
\begin{align}
\rateC&=1-\frac{(1-\fraction)\frac{\edges}{K}+\sum_{\ell=1}^{\nu\frac{\edges}{K}}\rows^{(\ell)}}{\n} =1-\frac{(1-\fraction)\frac{\edges}{K}+\nu\frac{\edges}{K}\rowavg}{\edges/J} =\rate_0-\fraction(1-\rate_0)(\rowavg-1).
\end{align}
\end{IEEEproof}

Note that the second term in \eqref{rate2} accounts for the rate loss at GC nodes. When the component codes are linear block codes with minimum distance $\dist$, we obtain the following bounds on $\rateC$:
\lemma{
If all component codes in the $\regC$ ensemble are linear block codes with minimum distance $\dist>2$, then
\begin{align}\label{converse}
\rateC\leq \rate_0-\fraction(1-\rate_0)\log_2\left(\frac{1}{2}\sum_{q=0}^{\lfloor \frac{\dist-1}{2}\rfloor}\binom {K} {q}\right).
\end{align}
Furthermore, there exists a set of linear block codes to be used as component codes such that
\begin{align}\label{achiv}
\rateC\geq \rate_0-\fraction(1-\rate_0)\left\lceil \log_2\left(\frac{1}{2}+\frac{1}{2}\sum_{q=0}^{\dist-2}\binom{K-1}{q}\right)\right\rceil.
\end{align}\label{lemma2}
Here, we use $\lceil \cdot\rceil$ and $\lfloor \cdot\rfloor$ to denote the ceiling and floor functions, respectively. The two bounds coincide, for example, when $\dist=3$ and $K=2^z-1$, where $z\in\mathbb{Z}_+$. 
}
\begin{IEEEproof}
First, the condition $\dist>2$ is required to differentiate between the rate loss at SPC nodes, which are block codes with minimum distance 2, and at GC nodes.  We start by proving the converse bound in \eqref{converse}. By the sphere-packing bound [15, Theorem 12, p.531],  any  component code with blocklength $K$ and minimum distance $\dist$ must satisfy
\begin{align}
2^{K-\rows}\leq \frac{2^K}{\sum_{q=0}^{\lfloor \frac{\dist-1}{2}\rfloor}\binom {K} {q}},
\end{align}
where $\rows$ is the number of rows in the parity-check matrix. Here we consider non redundant parity check matrices (i.e. $K- \rows$ is exactly the information dimension of the code). This implies that the term $(\rowavg-1)$ in \eqref{rate2} is bounded by 
\begin{align}\label{Hbound}
\rowavg-1\geq \log_2\left(\frac{1}{2}\sum_{q=0}^{\lfloor \frac{\dist-1}{2}\rfloor}\binom {K} {q}\right),
\end{align}
which proves \eqref{converse}. Regarding the achievable bound in \eqref{achiv}, the Varshamov Bound \cite[Theorem 2.9.3]{Huffman03} guarantees the existence of a linear component code with blocklength $K$ and minimum distance at least $\dist$ if 
\begin{align}
2^{K-\rows}\geq 2^{K-\left\lceil \log_2\left(1+\sum_{q=0}^{\dist-2}\binom{K-1}{q}\right)\right\rceil}.
\end{align}
If the above condition is satisfied, then there exists a set of linear block codes to be used as component codes with blocklength $K$ and minimum distance at least $\dist$ such that
\begin{align}\label{Vkavg}
\rowavg-1&\leq\left\lceil \log_2\left(\frac{1}{2}+\frac{1}{2}\sum_{q=0}^{\dist-2}\binom{K-1}{q}\right)\right\rceil,
\end{align}
which proves \eqref{achiv}.

Finally, if we substitute $\dist=3$ and $K=2^z-1$ for some $z\in\mathbb{Z}_+$ into \eqref{converse} and \eqref{achiv}, a straightforward computation shows that the converse bound in \eqref{converse} can be simplified to
\begin{align}\label{converse2}
\rateC\leq \rate_0-\fraction(1-\rate_0)(z-1),
\end{align}
and, likewise, the achievable bound in \eqref{achiv} simplifies to
\begin{align}\label{converse2}
\rateC\geq \rate_0-\fraction(1-\rate_0)(z-1).
\end{align}
\end{IEEEproof}

\subsection{Growth rate of the weight distribution of the $\regC$ ensemble}

A useful tool for analysis and design of LDPC codes and their generalizations is the asymptotic exponent of the weight distribution. The growth rate of the weight distribution was introduced in \cite{Gallager63} to show that the minimum distance of a randomly-generated regular LDPC code with variable nodes of degree of at least three is a linear function of the codeword length with high probability. The growth rate of the weight distribution for a class of doubly generalized LDPC (D-GLDPC) codes was introduced in \cite{Paolini13}. The $\regC$ GLDPC code ensemble can be seen as a particular instance of the codes analyzed in that work. The \emph{weight spectral shape} of the  $\regC$ ensemble captures the behavior of codewords whose weight is linear in the block length $\n$ and is defined by
\begin{align}\label{weight}
G(\alpha)\triangleq \lim_{\n\rightarrow\infty}\frac{1}{\n}\log \mathbb{E}_{\regC}[X_{\alpha \n}]   
\end{align}
for $\alpha>0$, where $X_{w}$ denotes the number of codewords of weight-$w$ of a randomly chosen  code in the $\regC$ code ensemble. This limit assumes the inclusion of only those positive integers for which $\alpha \n\in\mathbb{Z}$. We define the \emph{critical exponent codeword weight ratio} as $\hat{\alpha}\triangleq \inf\{\alpha\geq0 | G(\alpha)\geq 0\}$. If $\hat{\alpha}>0$, then the code's minimum distance asymptotically grows as $\mathcal{O}(\hat{\alpha}\n)$ and the ensemble is said to have good growth rate behavior. If $\hat{\alpha}=0$, then the minimum distance of the code may still grow with the block length $\n$ but at a slower rate, e.g., as $\mathcal{O}(\log(\n))$.
\lemma{
\label{lemma3}
If all component codes in the $\regC$ ensemble are linear block codes with minimum distance $\dist>2$, then $\hat{\alpha}>0$ for $J>2$. For $J=2$, $\hat{\alpha}>0$ if and only if
\begin{align}\label{fraction}
\fraction>\frac{K-2}{K-1}\triangleq\hat{\fraction}.
\end{align}
Otherwise, $\hat{\alpha}=0$.}
\begin{IEEEproof}
The lemma follows directly by particularizing the results in \cite{Paolini13} [Section II] to the $\regC$ ensemble.
\end{IEEEproof}

\begin{algorithm}[b]
\begin{algorithmic}
\STATE
\STATE
Remove from the Tanner graph of the GLDPC code all variable nodes with indexes in $\Gamma_{\y}$.
\STATE
Construct $\Psi$, the index set of check nodes that correspond to either degree-one SPC nodes or GC nodes of degree less  or equal to $\dist-1$.
\REPEAT 
\STATE
1) Select at random a member of $\Psi$.
\STATE
2) Remove from the Tanner  graph the check node with the index drawn in Step 1). Further, remove all connected variable nodes, and all attached edges.
\STATE
3) Update $\Psi$.
\UNTIL{All variable nodes have been removed (successful decoding) or $\Psi=\emptyset$ (decoding failure).}
\end{algorithmic}
\caption{BD-PD}\label{BD-PD}
\end{algorithm}

\section{Probabilistic Peeling Decoding over the BEC}\label{Peeling}

Suppose we use a random sample of the $\regC$ ensemble to transmit over a BEC($\ep$). For this channel, each of the $\n$ coded bits is erased with probability $\ep$. Without loss of generality, we assume that the all-zero codeword is transmitted,  hence the received vector $\y$ belongs to the set $\{0,?\}^\n$, where $?$ denotes an erasure. Let $\Gamma_{\y}\subseteq\{1,\ldots,\n\}$ be the index set of the bits correctly received, namely $y_i = 0$ for all $i\in\Gamma_{\y}$.  Decoding will be performed using a generalization of the PD algorithm \cite{Luby01} similar to that proposed for GLDPC codes in \cite{Olmos15}. The final formulation of the decoding algorithm depends on the decoding capabilities we assume at GC nodes. For instance, if we assume BD decoding at component codes, then the generalized PD algorithm, denoted as BD-PD, proceeds as described in Algorithm \ref{BD-PD}.

BD-PD is a suboptimal decoding method that considers decodable all GC nodes up to degree $\dist-1$ \cite{Jian2012,Miladinovic08}. However, it ignores the fact that any component code will be able to decode a certain fraction of erasure patterns of weight equal to or greater than $\dist$. As already reported in various works, e.g.,  \cite{Lentmaier10GLDPC, Olmos15}, the GLPDC code performance dramatically improves if we consider ML decoding at GC nodes. In principle, to consider ML decoding at GC nodes, we have to specify a full list of decodable erasure patterns and, label each of the incoming edges at every GC node to differentiate between decodable and non-decodable GC nodes.  As shown in \cite{Olmos15}, incorporating this labelling into the asymptotic analysis requires the use of multi-edge type DDs. 

\begin{figure}[t!]
\begin{tabular}{cc}
\hspace{3cm}\includegraphics{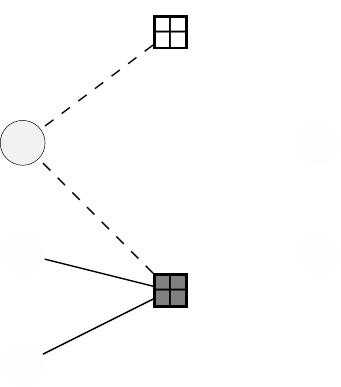} & \hspace{1cm}\includegraphics{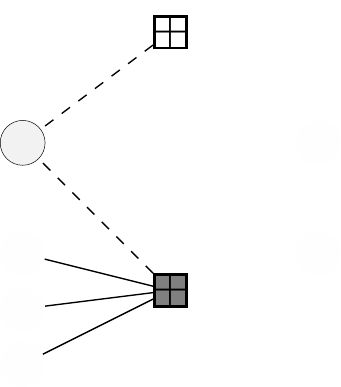}\\
(a) & (b)
\end{tabular}
\caption{We illustrate one iteration of the P-PD algorithm. Assuming GC nodes with $\dist=3$, in (a) right after dashed edges are removed, the remaining GC node (gray shadowed) becomes degree-2 and thus it will be considered decodable in future iterations.  In (b), after the GC node becomes degree-3, a sample from Bernoulli Random Variable with success probability equal to $p_3$ is drawn. If the sample is a success, we tagged the GC node as decodable for future iterations. Otherwise, it is tagged as non-decodable and only after the node losses any additional edge the tag can be reverted to decodable.}\label{P_PDprocess}
\end{figure}

\begin{algorithm}[htb]
\begin{algorithmic}
\STATE
\STATE
Remove from the Tanner graph of the GLDPC code all variable nodes with indexes in $\Gamma_{\y}$.
\FOR {all GC nodes}
\STATE
If the GC has degree $w$, tag the check node as \emph{decodable} with probability $p_w$.
\ENDFOR
\STATE
Construct $\Psi$, the index set of check nodes corresponding to either degree-one SPC nodes or GC nodes tagged as \emph{decodable}.
\REPEAT 
\STATE
1) Select at random a member of $\Psi$.
\STATE
2) Remove from the  Tanner \color{black} graph the check node with the index drawn in Step 1). Further remove all connected variable nodes and all attached edges.
\STATE 
3) \FOR {every \color{black}non-decodable \color{black} GC node that has lost one or more edges in the current iteration}
\STATE  If the GC has degree $w$, draw a sample of a Bernoulli distribution with success probability $p_w$. If the sample is a success, tag the check node as \emph{decodable}.
\ENDFOR
\STATE
4) Update $\Psi$.
\UNTIL{All variable nodes have been removed (successful decoding) or $\Psi=\emptyset$ (decoding failure).}
\end{algorithmic}
\caption{P-PD}\label{P-PD}
\end{algorithm}

In order to incorporate beyond-BD decoding at GC nodes into our analysis, and at the same time maintain a formulation compatible with the random definition of the $\regC$ ensemble, we will further constrain the family of component codes to be used at degree-$K$ GC nodes. More specifically, we assume that the fraction of ML-decodable weight-$w$ erasure patterns at every GC node is given by some $p_w\in[0,1], w=1,\ldots,K$. Thus, the family of component codes under analysis is the family of blocklength-$K$ linear block codes with minimum distance $\dist$ and with decoding profile described by the vector $\mathbf{p}=  ( \color{black} p_1, \ldots, p_K  ) \color{black}$. Note that if the minimum distance of the component code is $\dist$, then $p_w=1$ for $w\leq \dist-1$.   The bounds on $\rateC$, predicted in Lemma \ref{lemma2}, could in principle be refined according to $\mathbf{p}$. While this is an interesting open question, we will later show that the bounds are tight in certain scenarios and there is little room for refinement. 

By exploiting the fact that incoming edges at every GC node are  assigned to each position of the component code uniformly at random, we can incorporate ML-decoded GC nodes into the PD as shown in Algorithm \ref{P-PD}, denoted as \emph{probabilistic PD (P-PD)}. \textcolor{black}{Observe that the key P-PD feature is to tag GC check nodes as decodable with probabilities given by $\mathbf{p}$ only when they lose one or more edges, which may happen either at the initialization or after a connected variable is removed. If only one decodable check node is removed per iteration, after every P-PD iteration only a few GC nodes can change its state (from non-decodable to decodable). See Fig. \ref{P_PDprocess} for an explanatory diagram. Thus, at every iteration, P-PD emulates the ML decoding operation of a degree-$w$ GC node by drawing the decoding capability according to a Bernoulli distribution with parameter $p_w$, $w\in\{1,\ldots,K\}$. Note that P-PD is a procedure that allows for simpler analysis rather than a practical decoding algorithm.} Further, note that we recover the bounded distance PD (BD-PD) algorithm from P-PD if we set $p_w=0$ for $w\geq \dist$ and $p_w=1$ otherwise.

\subsection{Comparing the P-PD and ML-PD performances by Monte Carlo simulation}

If we select a specific component code, we can compare the simulation performance of the $\regC$ ensemble for the corresponding parameters under P-PD with that of the practical GLDPC codes with GC nodes that are decoded via ML, using the actual parity-check matrix of the component codes. We refer to this latter case as ML-PD. 

\begin{figure}[t]
\begin{tabular}{cc}
\hspace{-1cm}\includegraphics{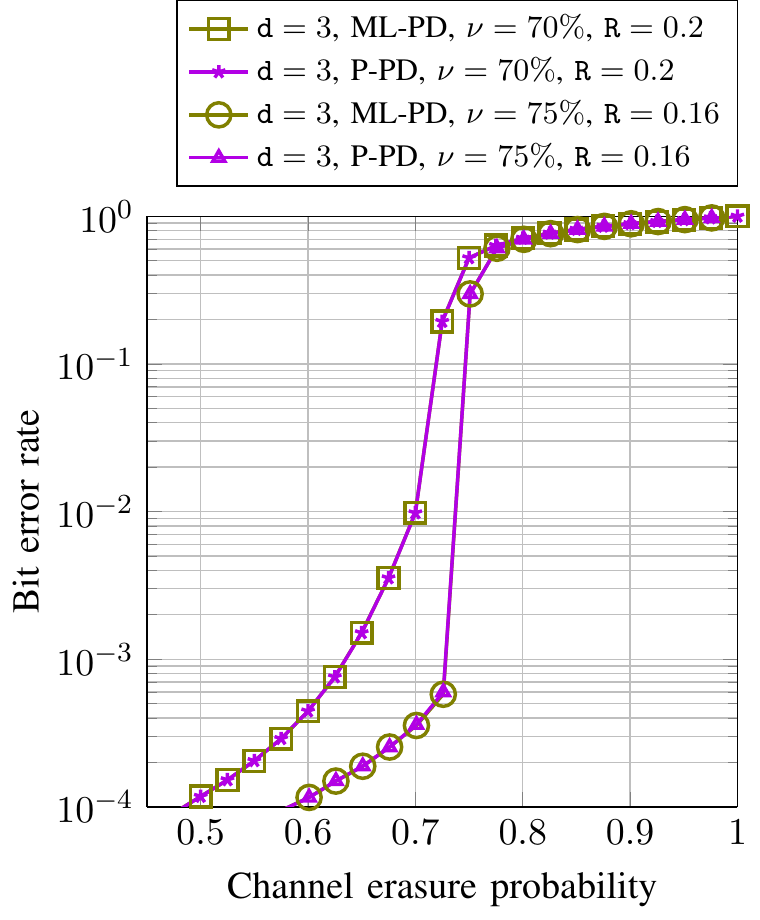} & \includegraphics{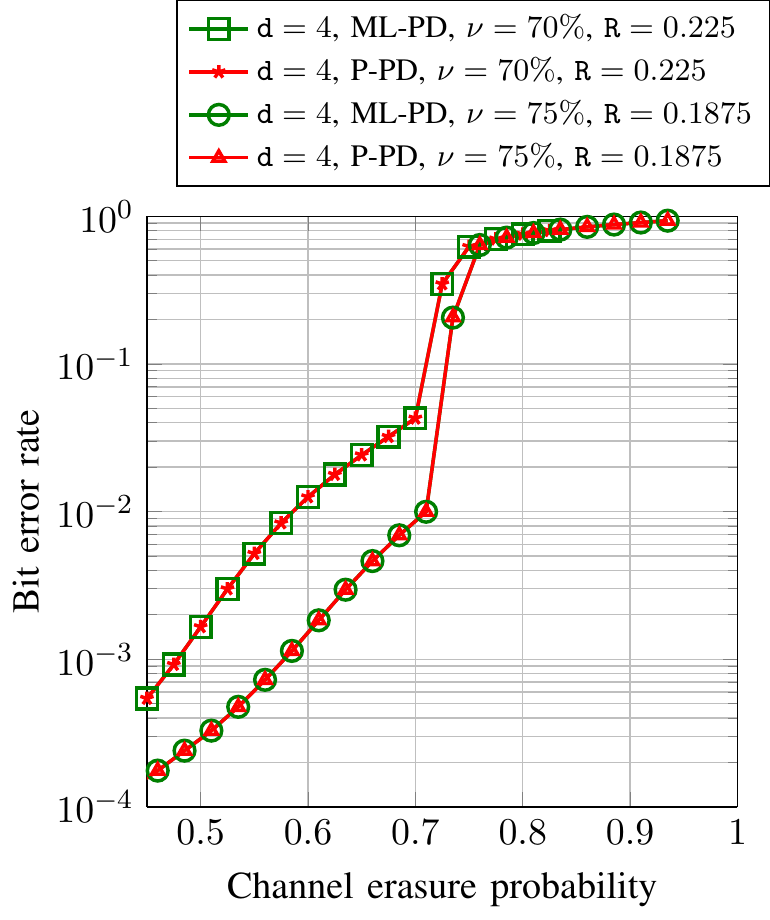}\\
(a) & (b)
\end{tabular}
\caption{In Fig. \ref{sims_26_28} (a), we plot the BER as a function of the channel erasure probability for a $(2,6)$ base DD and a rate-$1/2$ Hamming $(6,3)$ linear block code as component code. In Fig. \ref{sims_26_28} (b), we plot the BER as a function of the channel erasure probability for a $(2,8)$ base DD and a rate-$1/2$ $(8,4)$ Hamming component code. Results have been averaged over 10 generated samples from the $\regC$ ensemble with a blocklength of $\n=10000$ bits.}\label{sims_26_28}
\end{figure}

More precisely, for a given finite blocklength $\n$, fixed $\fraction\in[0,1]$, and base DD, we generate a member of the $\regC$ ensemble as follows:
\begin{enumerate}
\item Generate at random a Tanner graph according to the $(J,K)$ base DD. Then, select at random a fraction $\fraction$ of check nodes to be used as GC nodes. Overall, the graph contains $\n$ variable nodes, $\fraction\edges/K$ GC nodes and $(1-\fraction)\edges/K$ SPC nodes. 
\item For each of the $\fraction\edges/K$ GC nodes, we generate uniformly at random a permutation of the set $\{1,2,\ldots,K\}$, which is used to associate each of the incoming edges to the GC node to a position in the component code.
\end{enumerate}
We estimate by Monte Carlo simulation the bit error rate (BER) over the BEC achieved by both P-PD,  which follows Algorithm \ref{P-PD}, and ML-PD, which uses a look-up table of decodable erasure patterns. In Fig. \ref{sims_26_28} (a), we plot the BER as a function of the channel erasure probability of P-PD and ML-PD for a $(2,6)$-regular base DD with a rate-$1/2$ Hamming $(6,3)$ linear block code as component code. In Fig. \ref{sims_26_28} (b), we plot the same quantities for a $(2,8)$-regular base DD using a rate-$1/2$ $(8,4)$ Hamming component code. Results have been averaged over 10 generated samples from the $\regC$ ensemble. Observe the perfect match between the BERs for P-PD and ML-PD in all cases. This illustrates that we are not sacrificing accuracy with the probabilistic description of the decoder, as long as GLDPC codes are generated as described above. 

\section{Asymptotic analysis}\label{section_four}
The P-PD decoder yields a sequence of residual graphs by sequentially removing degree-one SPC nodes and decodable GC nodes from the GLDPC Tanner graph. Our next goal is to predict the asymptotic behaviour of the $\regC$ ensemble under P-PD by extending the methodology proposed in \cite{Luby01} to analyze the asymptotic behavior of LDPC ensembles under PD. In \cite{Luby01}, it is shown that if we apply the PD to elements of an LDPC ensemble, then the expected DD of the sequence of residual graphs can be described as the solution of a set of differential equations. Furthermore, the deviation of the process w.r.t. the expected evolution decreases exponentially fast with the LDPC blocklength. This analysis is based on a result on the evolution of Markov processes due to Wormald \cite{Wormald}. The proof that the GLDPC asymptotic graph evolution under P-PD can be predicted using the same result is given in Appendix \ref{app0}.  In this section, we introduce the notation used to characterize the DDs of the residual Tanner graphs of GLDPC ensembles with P-PD decoding and then present the system of differential equations that describes the asymptotic GLDPC graph evolution. In order to characterize the DDs of the residual Tanner graphs of GLDPC ensembles is to \nV{augment} the DD notation introduced in Section \ref{codes} to differentiate between GC nodes that have been tagged as decodable and those tagged as non-decodable. In order to simplify the formulation, we restrict ourselves to the case $p_w=0$ for $w\geq\dist+2$, i.e., we consider component codes can only decode a certain fraction of erasure patterns of degrees $\dist$ and $\dist+1$ and all erasure patterns of degree below $\dist$. This may not be an strong assumption. After exhaustive search of short linear block component codes (blocklengths up to 15 bits), we have not found any component code with $p_w>0$ for $w\geq\dist+2$. In any case, the analysis provided here directly generalizes to  any arbitrary $p_{w}$.

As introduced in Section \ref{codes}, any edge adjacent to a degree $i$ variable node is said to have left degree $i$, $i=1, \ldots,J$. Similarly, any edge adjacent to a degree $j$ SPC (GC) node is said to have right SPC (GC) degree $j$, $j=1, \ldots,K$.  Given the residual graph at the $\ell$-th iteration of the P-PD algorithm, let $L^{(\ell)} _i$ denote the number of edges with left degree $i$ at iteration $\ell$. Similarly, let $R^{(\ell)} _{pj}$ denote the number of edges with right SPC degree $j$ and $R^{(\ell)} _{cj}$ denote the number of edges with right GC degree $j$ at iteration $\ell$. For $j\in\{\dist,\dist+1\}$, we split $R^{(\ell)} _{cj}$ into two terms, $\hat{R}_{cj}^{(\ell)}$ and $\bar{R}_{cj}^{(\ell)}$, where $\hat{R}^{(\ell)} _{cj}$, $j\in\{\dist,\dist+1\}$ denotes the number of edges with right GC degree $j$ connected to GC nodes tagged as \emph{decodable}, and $\bar{R}^{(\ell)} _{cj}$ denotes the number of edges with right GC degree $j$ connected to GC nodes tagged as \emph{not-decodable}. Clearly, we have $R^{(\ell)} _{cj} = \hat{R}^{(\ell)} _{cj} +  \bar{R}^{(\ell)} _{cj}, j = \dist, \dist+1$. Recall that $\edges$ denotes the number of edges in the original GLPDC graph.


In the following theorem, we make use of Wormald's theorem \cite{Wormald} to show that the DD of the sequence of residual graphs during P-PD of a specific instance of the $\regC$ ensemble converges  to a function that can be computed by solving a set of deterministic differential equations. More specifically, for any element $Z^{(\ell)}\in\{L_i^{(\ell)},R_{pj}^{(\ell)},R_{cj}^{(\ell)}\}_{\substack{i=1,\ldots,J\\j=1,\ldots,K}}$ there exists a constant $\xi$ such that
\begin{align}\label{convergence}
P\left(\left| Z^{(\ell)}/\edges-z^{(\ell/\edges)}\right|>\xi\edges^{-\frac{1}{6}}\right)=\mathcal{O}\left(\text{e}^{-\sqrt{\edges}}\right),
\end{align}
where $z^{(\ell/\edges)}$ is the solution of a set of differential equations for that element of the DD, and  $\mathcal{O}\left(\text{e}^{-\sqrt{\edges}}\right)$ summarizes terms of order $\text{e}^{-\sqrt{\edges}}$. See Appendix \ref{app0} for more details. In the following, we use the notation  $Z^{(\ell)}/\edges\rightarrow  z^{(\ell/\edges)}$ to describe  convergence  in the sense of \eqref{convergence}.

\theorem{\label{ThP-PD}
Consider a BEC with erasure probability $\epsilon$ and assume we use elements of the  $\mathcal{C}_{\vlambda,\vrho_p,\vrho_c,\fraction}$ code ensemble for transmission. If we use P-PD  with parameters $(\dist,p_{\dist},p_{\dist+1})$, then the DD of the residual graph at iteration $\ell$ converges  to
\begin{align}\label{ex_g_l}
L_{i}^{(\ell)}/\edges&\rightarrow  l_{i}^{(\tau)}, ~~i\in\{1,\ldots,J\}\\\label{ex_g_r1}
R_{pj}^{(\ell)}/\edges&\rightarrow  r_{pj}^{(\tau)}, ~~ j\in\{1,\ldots,K\}\\\label{ex_g_r2}
R_{cj}^{(\ell)}/\edges&\rightarrow r_{cj}^{(\tau)}, ~~ j \in\{1, \dots, K\}\text{ and } j\notin\{\dist,\dist+1\}\\\label{ex_g_r3}
\hat{R}_{cj}^{(\ell)}/\edges&\rightarrow \hat{r}_{cj}^{(\tau)}, ~~j\in\{\dist,\dist+1\} \\\label{ex_g_r4}
\bar{R}_{cj}^{(\ell)}/\edges&\rightarrow \bar{r}_{cj}^{(\tau)} , ~~j\in\{\dist,\dist+1\}
\end{align}
where $l_{i}^{(\tau)}$, $r_{pj}^{(\tau)}$, $r_{cj}^{(\tau)}$, $\hat{r}_{cj}^{(\tau)}$, $\bar{r}_{cj}^{(\tau)}$, and $\tau=\frac{\ell}{\edges}\in[0,\sum_{i=1}^{J}l^{(\tau)}_i/i]$ are the solutions to the following system of differential equations:

\begin{align}\label{diff_gldpc_l1} 
\frac{\text{d}l_{i}^{(\tau)}}{\text{d}\tau}   &=   - \frac{i l^{(\tau)} _i }{e^{(\tau)}}\left(P_{p1}^{(\tau)} + \sum_{w = 1}^{\dist+1} w P_{cw}^{(\tau)} \right), \\\label{diff_gldpc_rp1} 
\frac{\text{d}r_{pj}^{(\tau)}}{\text{d}\tau} &=  P_{p1}^{(\tau)} \left((r^{(\tau)}_{p(j+1)} - r^{(\tau)} _{pj} ) \frac {  j (a^{(\tau)} -1) }{e^{(\tau)}} -\mathbbm{I}[j=1]\right) \nonumber\\&+  \sum_{w = 1}^{\dist +1}P_{cw}^{(\tau)} (r^{(\tau)}_{p(j+1)} - r^{(\tau)} _{pj} ) \frac { j w (a^{(\tau)} -1) }{e^{(\tau)}}, \\\label{diff_gldpc_rc1} 
\frac{\text{d}r_{cj}^{(\tau)}}{\text{d}\tau} &= P_{p1}^{(\tau)} \left((r^{(\tau)}_{c(j+1)} - r^{(\tau)} _{cj} ) \frac {  j (a^{(\tau)} -1) }{e^{(\tau)}} \right)  \nonumber\\
&+ \sum_{w = 1}^{\dist +1 }P_{cw}^{(\tau)} \left((r^{(\tau)}_{c(j+1)} - r^{(\tau)} _{cj} ) \frac { j w (a^{(\tau)} -1) }{e^{(\tau)}}-w\mathbbm{I}[j=w]\right),   ~~j \notin\{\dist, \dist+1\} 
\end{align}

\begin{align}\label{diff_gldpc_rc_dd1} 
\frac{\text{d}\hat{r}_{cj}^{(\tau)}}{\text{d}\tau} &= P_{p1}^{(\tau)} \left((p_{j} \bar{r}^{(\tau)}_{c(j+1)}+\hat{r}^{(\tau)}_{c(j+1)} - \hat{r}^{(\tau)} _{cj} ) \frac {  j (a^{(\tau)} -1) }{e^{(\tau)}} \right)\nonumber\\
&+ \sum_{w = 1}^{j +1}P_{cw}^{(\tau)} \left((p_{j} \bar{r}^{(\tau)}_{c(j+1)}+\hat{r}^{(\tau)}_{c(j+1)}  - \hat{r}^{(\tau)} _{cj} ) \frac { j w(a^{(\tau)} -1) }{e^{(\tau)}}-w \mathbbm{I}[w=j]\right) , ~~j \in\{\dist, \dist+1\} \\\nonumber
\frac{\text{d}\bar{r}_{cj}^{(\tau)}}{\text{d}\tau}  &= P_{p1}^{(\tau)} \left(((1-p_{j}) \bar{r}^{(\tau)}_{c(j+1)} - \bar{r}^{(\tau)} _{cj} ) \frac {  j (a^{(\tau)} -1) }{e^{(\tau)}} \right) \\\label{diff_gldpc_rcbar}
&+\sum_{w = 1}^{j +1}P_{cw}^{(\tau)} \left(((1-p_{j}) \bar{r}^{(\tau)}_{c(j+1)}  - \bar{r}^{(\tau)} _{cj} ) \frac { j w (a^{(\tau)} -1) }{e^{(\tau)}} -w \mathbbm{I}[w=j] \right),    ~~j \in\{\dist, \dist+1\} 
\end{align}
In \eqref{diff_gldpc_l1}-\eqref{diff_gldpc_rcbar}, $\mathbb{I}[\cdot]$ denotes the indicator function, and
\begin{align}\label{defe}
 e^{(\tau)} &= \sum_{i=1}^{J}l^{(\tau)} _i = \sum_{j=1}^{K}[r^{(\tau)} _{pj}+r^{(\tau)} _{cj}] ,\\
 a^{(\tau)} &= \sum_{i} i l_i^{(\tau)} / e^{(\tau)}, \\\label{defp1}
 P_{p1}^{(\tau)} &=\frac{r_{p1}^{(\tau)}}{s^{(\tau)}}, \\\label{defpj}
 P_{cj}(\tau) &=\left\{\begin{array}{cc} \displaystyle \frac{r^{(\tau)} _{cj}/j}{s^{(\tau)}}  & j< \dist \\ \\ \displaystyle \frac{\hat{r}^{(\tau)} _{cj}/j}{s^{(\tau)}} & j\in\{\dist,\dist+1\} \end{array}\right.\\
  s^{(\tau)} &= r_{p1}^{(\tau)} +  \sum_{w =1}^{\dist-1} \frac{r^{(\tau)} _{cw}}{w} + \frac{\hat{r}^{(\tau)} _{c\dist}}{\dist} + \frac{\hat{r}^{(\tau)} _{c(\dist+1)}}{\dist+1}.
 \end{align}
The initial conditions of the system of differential equations \eqref{diff_gldpc_l1}-\eqref{diff_gldpc_rcbar} are given by 
\begin{align} \label{diff_gldpc_l_ini}
l_{i}^{(0)}  &= \epsilon \lambda_i, \\\label{diff_gldpc_ini_rp}
r_{pj}^{(0)}  &= \sum_{\alpha \geq j} \rho_{p\alpha} {\alpha-1 \choose j-1} \epsilon ^{j} (1-\epsilon)^{\alpha -j}, \\\label{diff_gldpc_ini_rc}
r_{cj}^{(0)} &= \sum_{\alpha \geq j} \rho_{c\alpha} {\alpha-1 \choose j-1} \epsilon ^{j} (1-\epsilon)^{\alpha -j}, \\\label{diff_gldpc_l_ini3}
\hat{r}_{c\nu}^{(0)} &= p_{\nu}r_{c\nu}^{(0)}, \\\label{diff_gldpc_l_ini4}
\bar{r}_{c\nu}^{(0)} &= (1-p_{\nu})r_{c\nu}^{(0)}
\end{align}
for $i=1,\ldots J$, $j=1,\ldots,K$, and $\nu = \dist, \dist+1$. 
}
 \begin{IEEEproof}
See  Appendix \ref{app0}.
\end{IEEEproof}

Using Theorem \ref{ThP-PD}, we can predict the P-PD threshold for the $\regC$ code ensemble by setting $\lambda_i=\mathbbm{I}[i=J]$ in \eqref{diff_gldpc_l_ini}, $\rho_{p\alpha}=(1-\fraction)\mathbbm{I}[\alpha=K]$ in \eqref{diff_gldpc_ini_rp}, and $\rho_{c\alpha}=\fraction\mathbbm{I}[\alpha=K]$ in \eqref{diff_gldpc_ini_rc}. We then  numerically search for the highest $\epsilon$ value for which the function $r^{(\tau)} _{p1}+\sum_{w =1}^{\dist-1}r^{(\tau)} _{cw}/w + \hat{r}^{(\tau)} _{c\dist}/\dist+\hat{r}^{(\tau)} _{c(\dist+1)}/(\dist+1)$ remains strictly positive for any $\tau\in[0,\sum_{i=1}^{J}l^{(\tau)}_i/i]$ such that $e^{(\tau)}>0$. 

\subsection{An upper bound on the iterative-decoding threshold }

For standard LDPC code ensembles, it is  known that the BP iterative decoding threshold is upper bounded by the so-called \emph{stability condition}  (SC) \cite{Rich-Sho-Urb-LDPC01}:
\begin{align}
\epsilon^* \leq \left[\lambda_2~\rho'(1)\right]^{-1},
\end{align}
where $\rho(x)$ is the right degree polynomial, $\rho'(1)$ its derivative at $x=1$ and $\lambda_2$ is the fraction of edges in the graph with left degree equal to 2.   In \cite{Paolini07}, Paolini, Fossorier, and Chiani extended the bound for GLDPC code ensembles by performing a Taylor expansion of the asymptotic GLDPC EXIT function. In particular, they proved that if the GLDPC code ensemble only contains generalized component codes with $\dist\geq3$, then the iterative decoding threshold is upper bounded by
\begin{align}
\epsilon^* \leq \left[\lambda_2~\rho'_p(1)\right]^{-1},
\end{align}
where 
\begin{align}
\rho_p(x)  = \sum_{j\geq 2} \rho_{pj} x^{j-1},
\end{align}
and $\rho_{pj}$, as defined in Section \ref{codes}, is the fraction of edges in the GLDPC Tanner graph connected to degree-$j$ SPC nodes. For the $\regC$ ensemble with $J=2$, this bound simplifies to
\begin{align}\label{scgldpc}
\epsilon^* \leq \frac{1}{(K-1)(1-\fraction)},
\end{align}
while for $J>2$ this bound is non-informative (it is infinite) since $\lambda_2=0$.
\section{Analysis of the $\regC$ ensemble under P-PD}\label{SecV}

In this section, we study the asymptotic performance of the $\regC$ ensemble for different base DDs as we vary the fraction $\fraction$ of GC nodes in the graph. We use high rate  base DDs that correspond to regular LDPC code ensembles with variable degree equal to $J=2$. Further examples with $J>2$ are discussed in Sections \ref{irr} and \ref{doubly}. We summarize the parameter of the base DD considered here in  Table \ref{table0}. We denote by $\PDth$ the PD threshold of the base LDPC ensemble. Recall that $p_{w}=1$ for $w\leq \dist-1$ and $p_{w} = 0$ for $w \geq \dist +2$. 
In order to determine $p_{\dist}, p_{\dist+1}$, we performed an exhaustive search over the database \cite{Grassl:codetables,Grassl06}, which implements MAGMA \cite{Magma} to design block codes with the largest minimum distance. For every $K$, we search for the code with the largest minimum distance $\dist$, and we use the corresponding $p_\dist$ and $p_{\dist+1}$ parameters.
Like this, we ensure that there exists at least one linear block code that satisfies these requirements. We use this specific block code as the reference of a family of linear block codes with the same decoding capabilities. The values found are listed in Table \ref{table1} and used as a reference for a whole family of linear block codes. The corresponding reference block codes are listed in Appendix \ref{app1}. Note that despite having different blocklength and rate, many reference block codes share the same $p_{\dist}$, $p_{\dist+1}$ parameters. 


We construct $\regC$ ensembles by combining various base DDs with the component code families summarized in Table \ref{table1}. For each code ensemble, we compute the P-PD threshold $\th$ as a function of $\fraction$.


\begin{table}[t!]
\centering
\caption{Base DDs, their design rates and iterative decoding thresholds under PD}\label{table0}
\begin{tabular}{c|cccc}
Base DD   & $K$ &$\rate_0$ & $\PDth$ & Gap to capacity ($1-\rate_0-\epsilon_0$) \\
\hline
$(2,6)$-regular   & 6  & 2/3 & 0.206 & 0.127\\
$(2,7)$-regular  & 7 & 5/7 & 0.167  & 0.119 \\
$(2,8)$-regular  &  8 & 3/4  & 0.147 & 0.103 \\
$(2,15)$-regular  &  15 & 13/15  & 0.071  & 0.062\\
\end{tabular}
\end{table}

\begin{table}[h!]
\centering
\caption{Families of component linear block codes. }\label{table1}
\begin{tabular}{c|cccccc}
Code Family Index & blocklength $K$ & $\dist$ &  $p_{\dist}$ &  $p_{\dist+1}$ \\
\hline
I & 6 & 3 & 0.8 &  0 \\
II & 6 & 4 & 0.8 &  0  \\
III & 7 & 3 & 0.8 & 0 \\
IV & 7 & 4 & 0.8 & 0 \\
V & 8 & 4 & 0.8 &  0  \\
VI & 8  & 4 & 0.9143 &  0.5714  \\
VII & 8  & 5 & 0.9643 &  0.75  \\
VIII & 15   & 3 & 0.9231 &  0.6154  \\
IX & 15  & 4 &  0.9231 & 0.6154\\ 
\end{tabular}
\end{table}

\subsection{Results for $(2,6)$ and $(2,7)$ base DDs}

\begin{figure}[h!]
\centering
\includegraphics{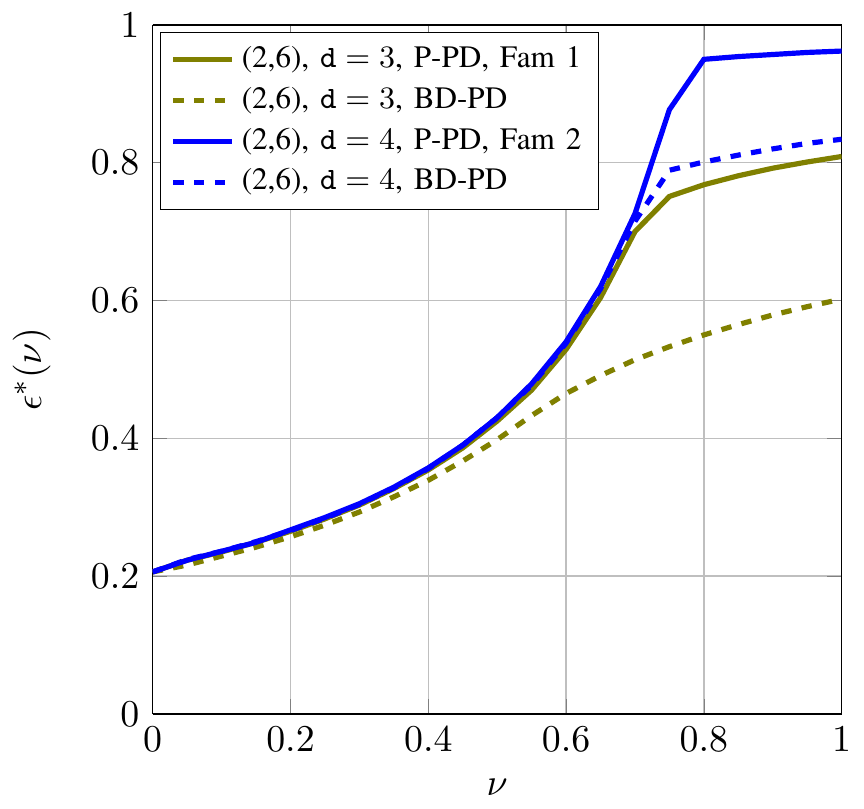}
\caption{P-PD and BD-PD thresholds  as a function of $\fraction$ for the $(2,6)$ base DD.}\label{GLDPCth}
\end{figure}

Fig. \ref{GLDPCth} shows the computed P-PD  threshold $\th$ of the $\regC$ ensemble for a base DD $(2,6)$-regular as a function of $\fraction$. We consider GC nodes with minimum distance $\dist$ equal to 3 and 4 and parameters given by Families I and II in Table \ref{table1}. We also include the BD-PD threshold, which only depends on the minimum distance $\dist$ of the component codes and can be computed by solving the system of differential equations in Theorem \ref{ThP-PD} by setting $p_{\dist}=p_{\dist+1}=0$. First of all, observe that the P-PD gains in threshold w.r.t. BD-PD are only significant for large values of $\fraction$. Furthermore, for both P-PD and BD-PD, using component codes with larger minimum distance ($\dist=4$ instead of $\dist=3$) pays off only for very large values of $\fraction$. 

Since increasing $\fraction$  also modifies the code rate $\rateC$ in \eqref{rate2}, the comparison in Fig. \ref{GLDPCth} can be misleading, as we cannot directly evaluate the distance to the channel capacity. In fact, not all values of $\fraction$ are achievable, since they would give rise to a negative rate $\rateC$. We overcome this issue by directly comparing the asymptotic threshold and code rate, both defined as parametric curves w.r.t. $\fraction$. Denote by $\th(\fraction)$ the threshold $\th$ as a function of $\fraction\in[0,1]$. From Fig. \ref{GLDPCth} we see that $\th(\fraction)$ is a continuous, strictly increasing function of $\fraction$ and that for $\fraction=0$ its value is equal to $\PDth$, the threshold of the base LDPC ensemble. The inverse of this function, which can be obtained numerically, is denoted by $\fraction(\th)$ and provides the minimum fraction of GC nodes in the graph  required to achieve an ensemble threshold at least $\th$. Given the function $\fraction(\th)$ described above, we  use Lemma \ref{lemma2} to determine bounds on $\rateC$ for a given targeted decoding threshold $\th$. More precisely, by using $\fraction(\th)$ in \eqref{converse}, we obtain a \emph{converse bound} on the coding rate required to achieve a P-PD decoding threshold equal to $\th$ using component codes with minimum distance $\dist$. Similarly, using $\fraction(\th)$ in \eqref{achiv}, we obtain an \emph{achievable bound} on the coding rate required to achieve a P-PD decoding threshold equal to $\th$ using linear component codes with minimum distance $\dist$. We proceed along the same lines to obtain bounds on the $\regC$ rate for the BD-PD thresholds. 

\begin{figure}[t!]
\begin{tabular}{cc}
\hspace{0cm}\includegraphics{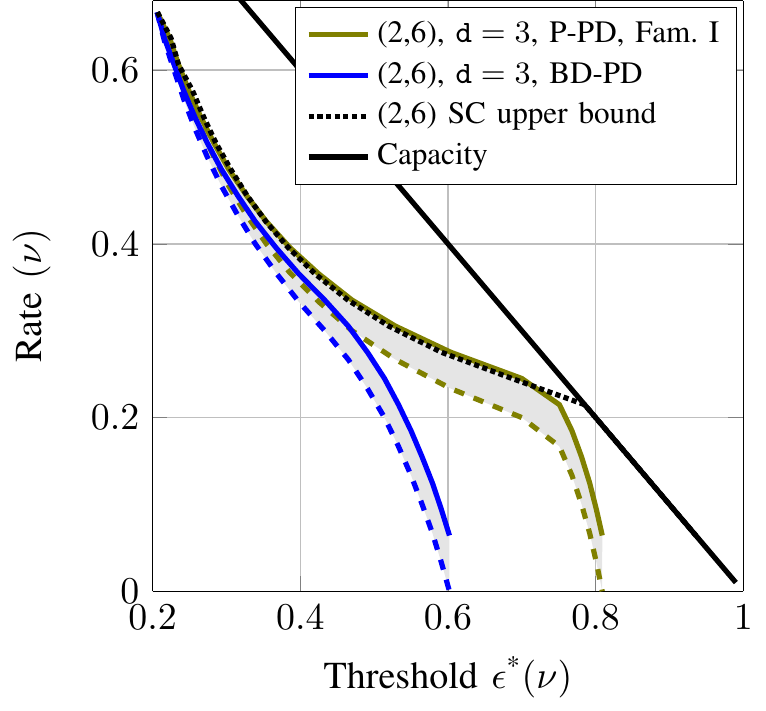} & \includegraphics{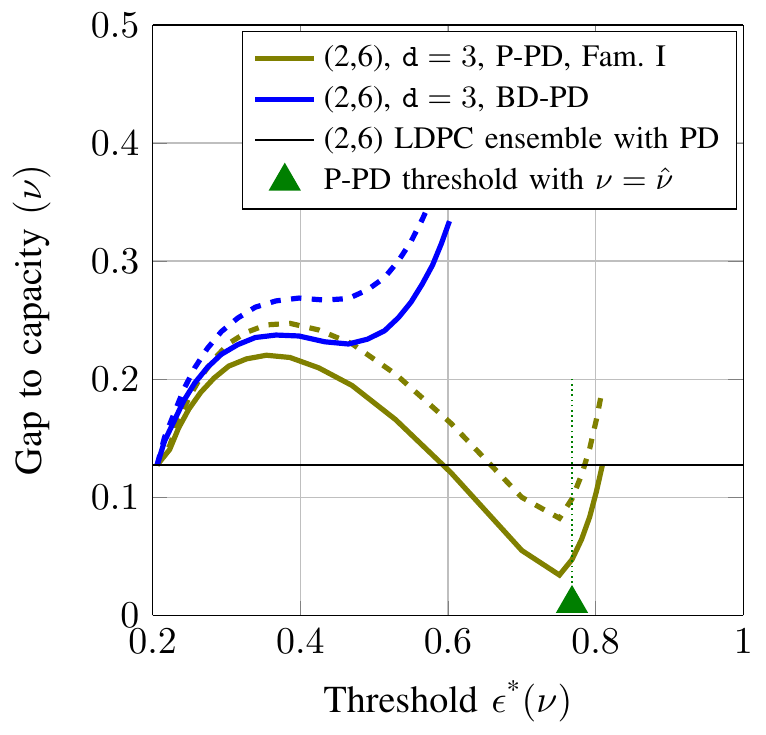}\\
(a) & (b)
\end{tabular}
	\caption{ In Fig. \ref{rate_threshold_261} (a), we plot the bounds on the $\regC$ coding rate in \eqref{converse} and \eqref{achiv} for the base DD $(2,6)$ and component codes of minimum distance $\dist=3$ as a function of the P-PD and BD-PD thresholds. In Fig. \ref{rate_threshold_261} (b), we show their gap to channel capacity. We also indicate the P-PD threshold for $\fraction=\hat{\fraction}$.}\label{rate_threshold_261}
\end{figure}

In Fig. \ref{rate_threshold_261} (a) we plot these bounds as a function of $\th$, both for P-PD and BD-PD, using Code Family I component codes with minimum distance $\dist=3$. We further include the SC upper bound in \eqref{scgldpc}. Observe that \eqref{scgldpc} coincides with the rate-threshold converse bound in \eqref{converse} up to $\fraction \approx 0.75$. Above $\fraction=0.8$, the SC bound exceeds channel capacity. 

In Fig. \ref{rate_threshold_261} (b), we show the gap to channel capacity computed for each case, and indicate the threshold $\epsilon^*(\hat{\fraction})$ with $\hat{\fraction}$ given in \eqref{fraction}. Since $\epsilon^*(\fraction)$ is monotonically increasing in $\fraction$, any configuration with threshold larger than $\epsilon^*(\hat{\fraction})$ has a minimum distance that grows linearly with the block length $\n$. Observe that the performance of both BD-PD and P-PD overlaps for coding rates close to the original rate of the base DD, i.e., for small values of $\fraction$. However, as $\epsilon^*(\fraction)$ increases, P-PD significantly outperforms BD-PD. Furthermore, there are values of $\fraction$ for which the gap to capacity of P-PD is smaller than that for the base LDPC ensemble under PD. For the $(2,6)$ base DD, the minimum gap to capacity of P-PD, measured using the achievable rate bound, is 0.0823 for a coding rate of 0.1667. For $\fraction=\hat{\fraction}$, the gap to capacity grows to 0.0987 but it is still below the base LDPC  gap to capacity, which is 0.1273 according to Table \ref{table0}. Thus, for $\fraction$ slightly above $\hat{\fraction}$ we are able to reduce the original gap to capacity and at the same time obtain a good ensemble from minimum distance point of view. Observe also that the region where the $\regC$ ensemble outperforms the base LDPC ensemble is very narrow,  and it does not include the case where all check nodes are GC nodes $(\nu=1)$. 


\begin{figure}[h!]
\centering\includegraphics{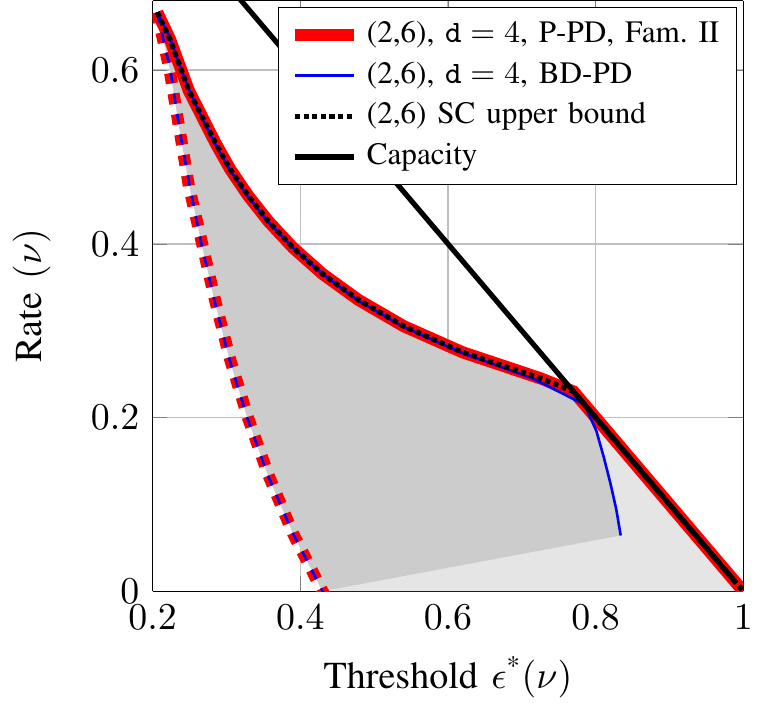}
\caption{Bounds on the $\regC$ coding rate in \eqref{converse} and \eqref{achiv} for a base DD $(2,6)$ and $\dist=4$ component codes as a function of the P-PD and BD-PD thresholds. }\label{rate_threshold_261_4}
\end{figure}

Fig. \ref{rate_threshold_261_4} reproduces the results for the Code Family II with minimum distance $\dist=4$. However, in this case the two bounds are loose and it is uncertain whether we can find an specific block component code in the family that is able to operate close to the converse bound. The P-PD converse bound now overlaps with the SC bound in the whole regime and,  for large $\epsilon^{*}(\fraction)$, it coincides with the capacity. Furthermore, the bounds for P-PD and BD-PD overlap in a large region despite the fact that P-PD using component codes from Family II resolves degree-$\dist$ erasure patterns with high probability ($0.8$).  

\begin{figure}[h!]
\begin{tabular}{cc}
\hspace{0cm}\includegraphics{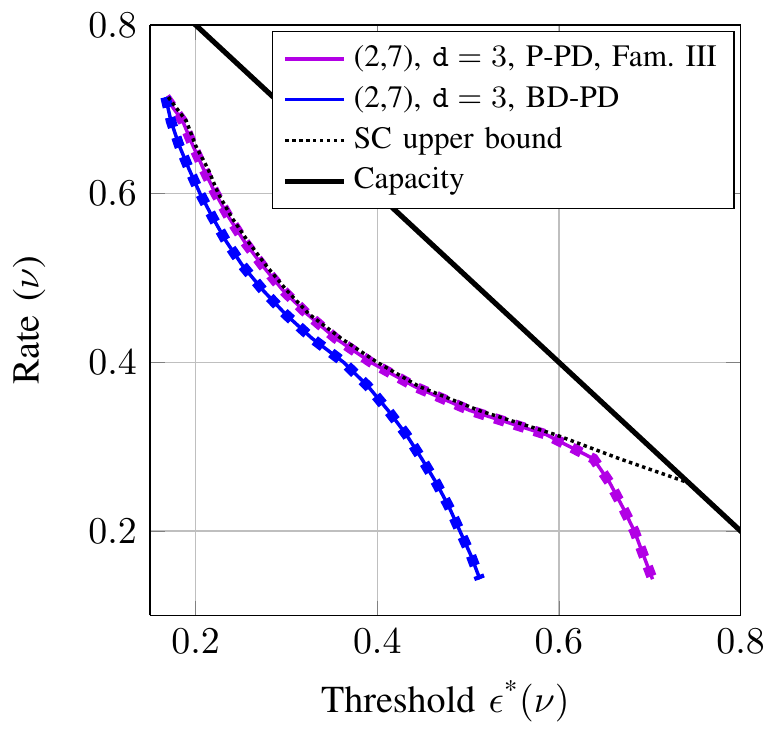} & \includegraphics{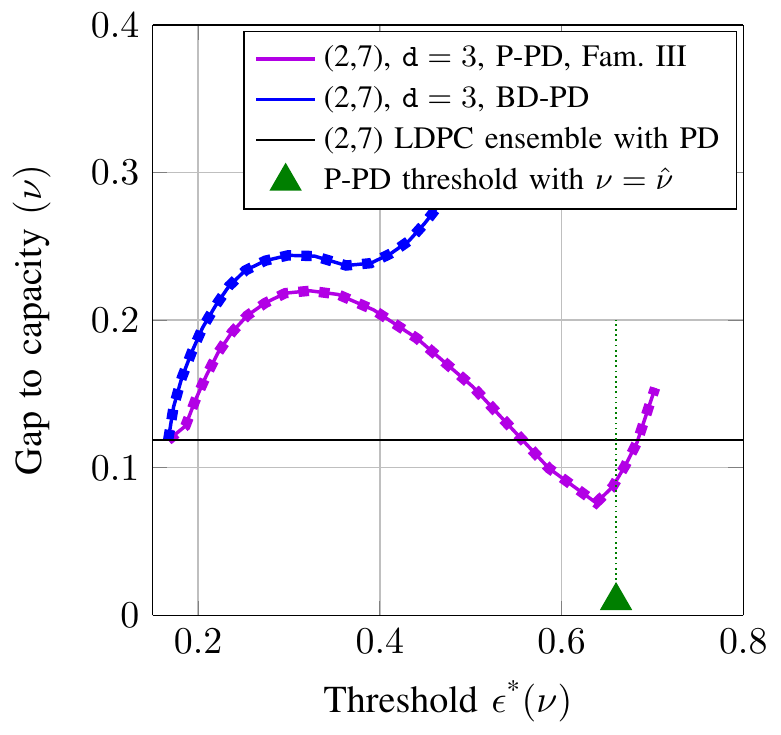}\\
(a) & (b)
\end{tabular}
\caption{ In (a), we plot the bounds on the $\regC$ coding rate in \eqref{converse} and \eqref{achiv} for a base DD $(2,7)$ as a function of the P-PD and BD-PD thresholds. Note that the bounds overlap in this case. In (b), we show the gap to channel capacity for each case. We also indicate the P-PD threshold for $\fraction=\hat{\fraction}$.}\label{rate_threshold_27}
\end{figure}

In Fig. \ref{rate_threshold_27} we show the asymptotic behaviour of the $\regC$ ensemble constructed using a $(2,7)$ base DD with $\dist=3$ component codes. As predicted by Lemma \ref{lemma2}, when using component codes of blocklength $K=7$ with minimum distance $\dist=3$, the converse and achievable bound on the $\regC$ coding rate coincide. Thus, the existence of a linear block component code that satisfies the properties of Code Family III and for which the  $\regC$ ensemble asymptotically achieves the results in Fig. \ref{rate_threshold_27} is guaranteed. Again, there is a region where the gap to capacity of P-PD can be reduced with respect to that of the base LDPC ensemble, which is roughly aligned with the point where the P-PD threshold separates from the SC upper bound in \eqref{scgldpc}.


\subsection{Results for higher-density base DDs}

\begin{figure}[h!]
\begin{tabular}{cc}
\hspace{0cm}\includegraphics{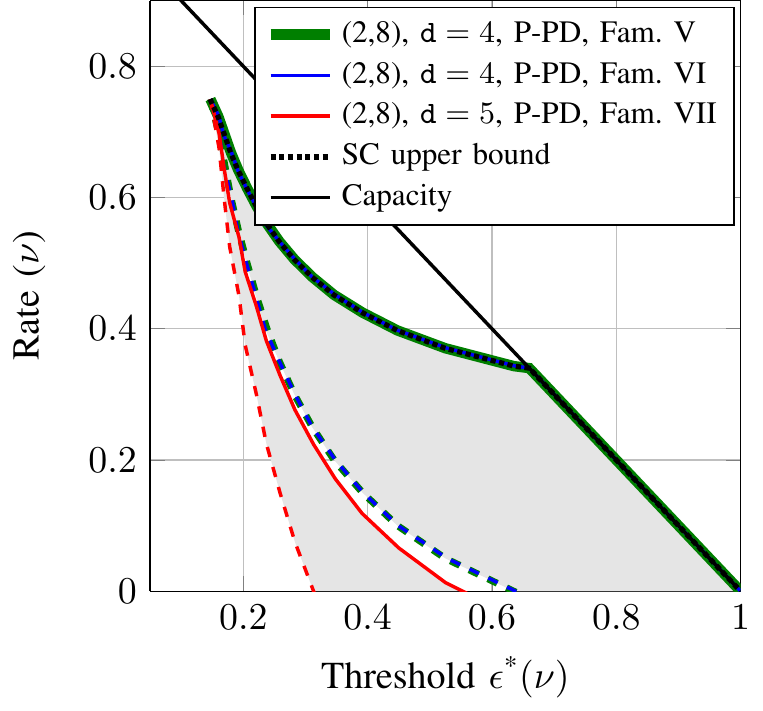} & \includegraphics{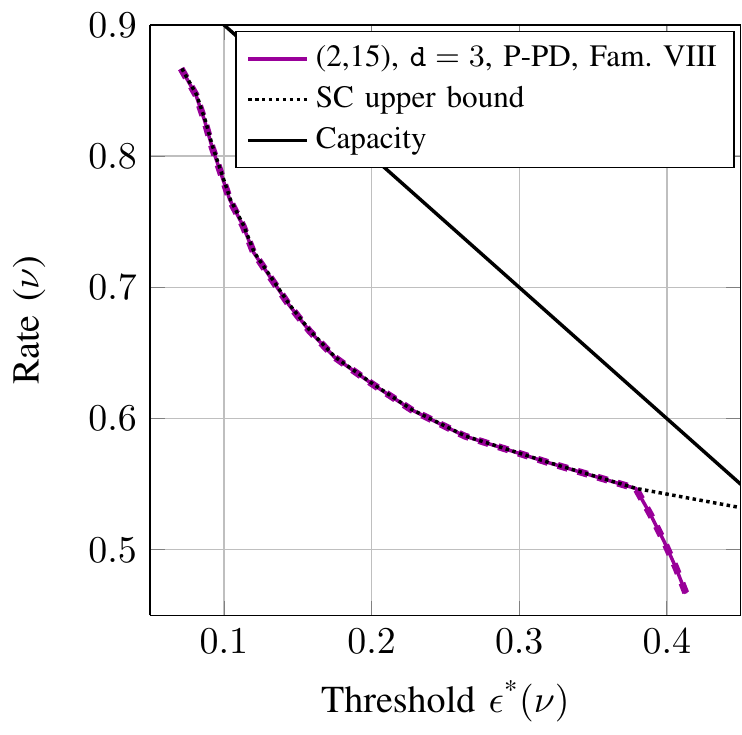}\\
(a) & (b)
\end{tabular}
\caption{ We plot the bounds on the $\regC$ coding rate in \eqref{converse} and \eqref{achiv} for a base DD $(2,8)$ (Fig. \ref{rate_threshold_higher} (a)) and $(2,15)$ (Fig. \ref{rate_threshold_higher} (b)) as a function of the P-PD threshold. }\label{rate_threshold_higher}
\end{figure}

We finish this section by extending the above results to base DDs with higher check degree and, thus higher ensemble density. In Fig. \ref{rate_threshold_higher}(a), we show the asymptotic behavior of the $\regC$ ensemble constructed using a $(2,8)$ base DD with component codes in Code Families V, VI and VII (See Table \ref{table1}). Observe first that the rate bounds for Code Families V and VI coincide, even though Code Family VI has better decoding capabilities. In both cases the bounds are loose, but we can still observe a significant improvement w.r.t. the Code Family VII, which has but very large ($\dist=5$) minimum distance and, hence, and small coding rate. This again illustrates the  trade-off between the threshold performance and the rate penalty induced by considering lower rate GC nodes.
 In Fig. \ref{rate_threshold_higher}(b), we consider a $(2,15)$ base DDs with a component code of Code Family VIII $(\dist =3)$. In this case, as predicted by lemma \ref{lemma2}, the bounds coincide and the gap to capacity is minimized at a coding rate $\rate\approx 0.54$ and threshold $\epsilon^*\approx0.379$, resulting in a gap capacity equal to $0.074$. This is slightly above the gap to capacity for the base LDPC ensemble ($0.062$). Also, at this point the GLDPC ensemble does not have linear growth of the minimum distance, since for this ensemble, $\epsilon^*(\hat{\fraction})=0.493$. 


\color{black}
\section{Selecting specific component codes}\label{MLPD}

By using the bounds on the $\regC$ code rate, we have been able to assess the performance of $\regC$ ensembles for a family of linear component codes. In certain scenarios the proposed bounds on the $\regC$ code rate provide meaningful design information about the asymptotic behavior of the ensemble. The natural question that arises at this point is whether we can find specific component codes within the family that outperform the achievable bound in \eqref{achiv}, reducing the gap to the rate converse bound in \eqref{converse}. In this section, we analyze the asymptotic performance of $\regC$ when component codes are chosen from the the list of reference linear block component codes summarized in Table \ref{tablecodes}. The construction of these linear block codes is detailed in \cite{Grassl:codetables}, and their generator matrix is given in Appendix \ref{app1}. We use the notation R-{I} to denote the reference linear block code of Code Family I.

\begin{table}[h!]
\centering
\caption{Reference component codes. The parameter $\rows$ describes the number of rows in the parity-check matrix. }\label{tablecodes}
\begin{tabular}{c|cccc}
Code index& Blocklength $K$  & $\rows$ & Rate & Code family in Table \ref{table1}  \\
\hline
R-I & 6 & 3  & 1/2&  I \\
R-II & 6 & 4  & 1/3&  II \\
R-III & 7 & 3 & 4/7& III  \\
R-IV & 7 & 4 & 3/7& IV  \\
R-V & 8 & 4  & 1/2 & V  \\
R-VI & 8 & 5  & 3/8 &  VI  \\
R-VII & 8  & 6 & 1/4 & VII \\
R-VIII & 15  & 4 & 11/15 & VIII   \\
R-IX & 15 & 5 & 2/3 & IX \\ 
\end{tabular}
\end{table}

\begin{figure}[h!]
\centering\includegraphics{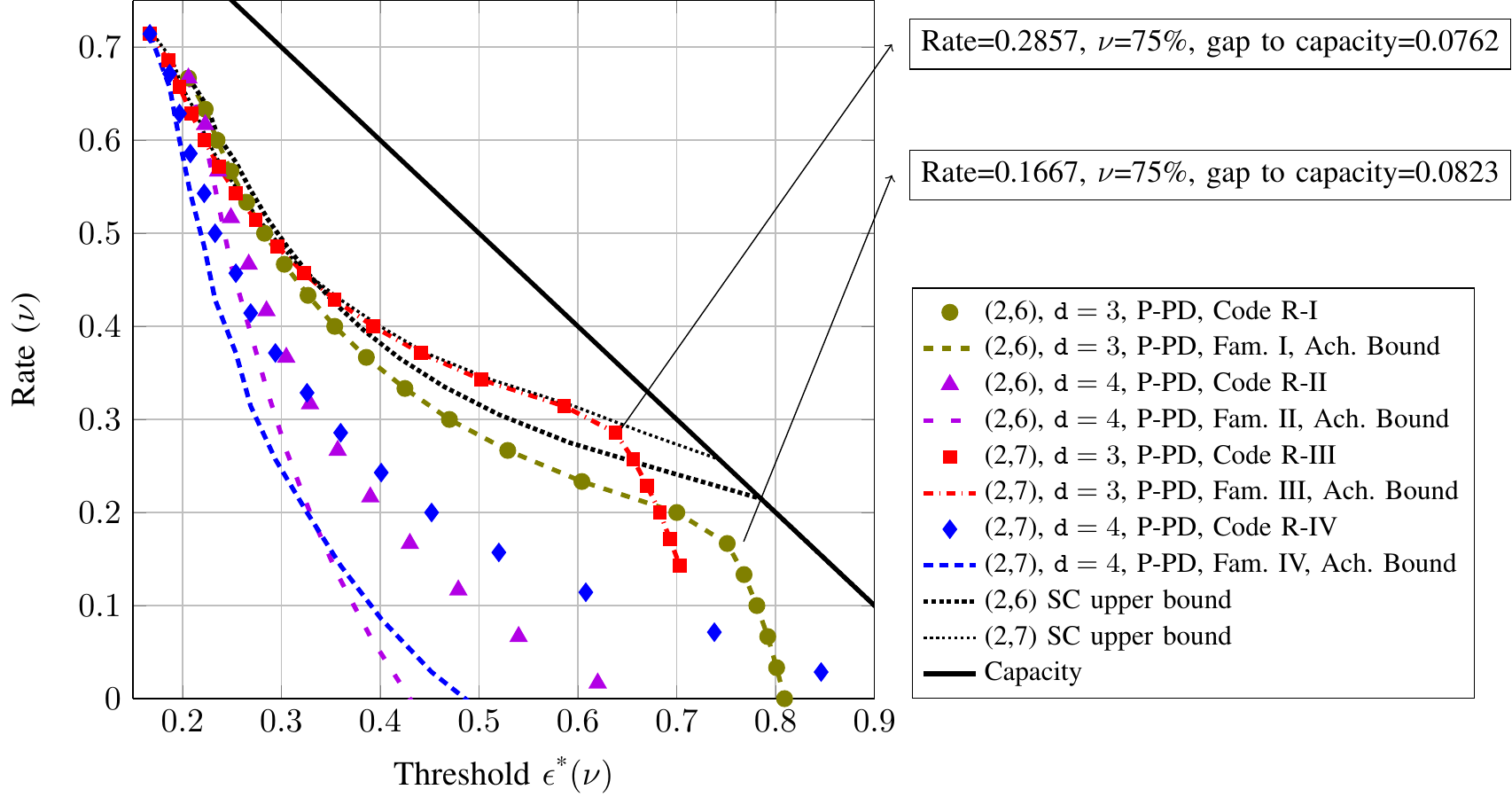}
\caption{$\regC$ coding rate and achievable bound in \eqref{achiv} for $(2,6)$ and $(2,7)$ base DDs and component codes from Table \ref{table1} and \ref{tablecodes} as a function of the P-PD decoding threshold.}\label{rate_threshold_261_code}
\end{figure}

\begin{figure}[h!]
\centering\includegraphics{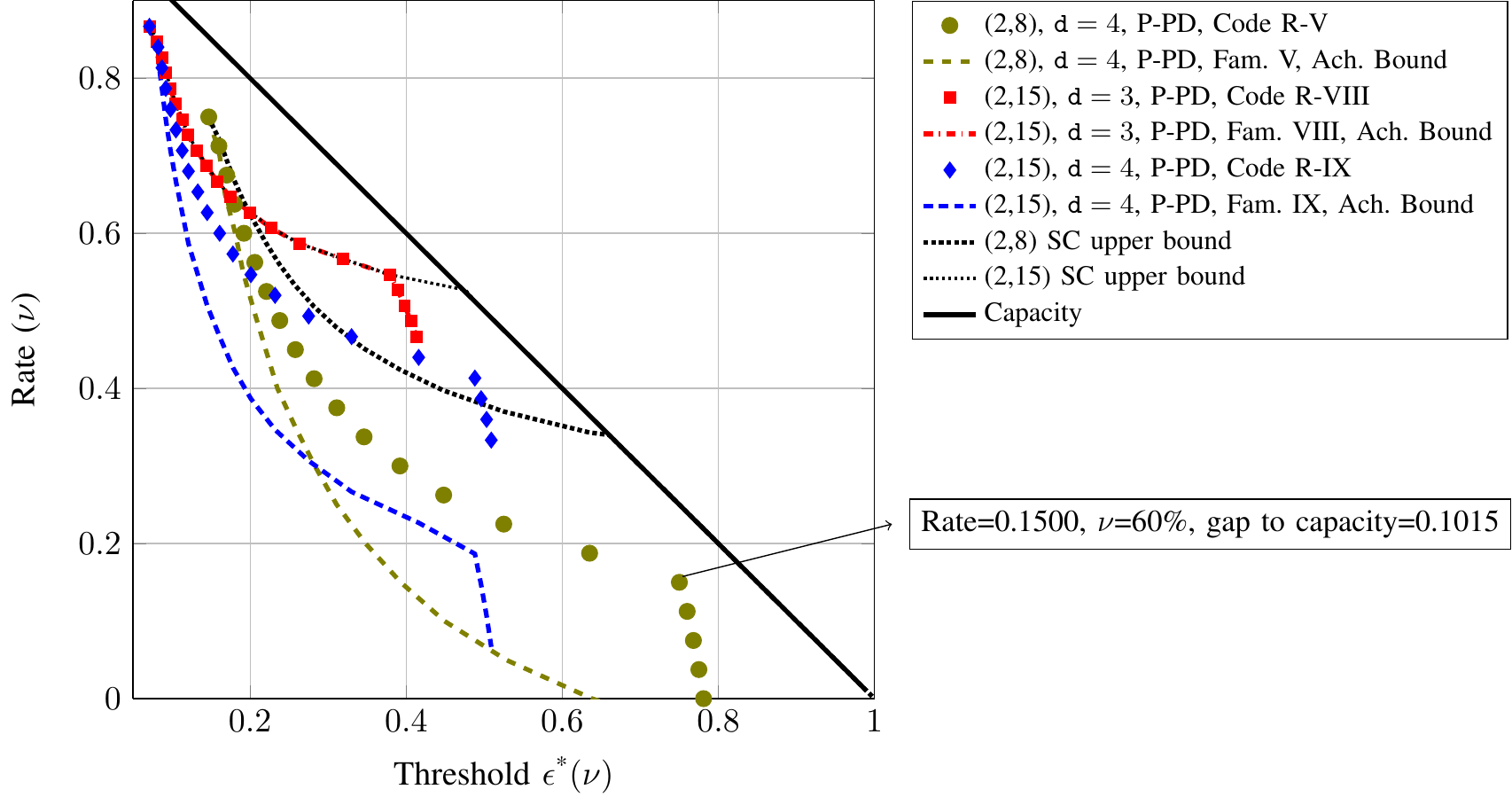}
\caption{ $\regC$ coding rate and achievable bound in \eqref{achiv} for $(2,8)$ and $(2,15)$ base DDs and component codes from Tables \ref{table1} and \ref{tablecodes} as a function of the P-PD decoding threshold.}\label{rate_threshold_281_4}
\end{figure}

Once we fix a particular class of component codes to be used at GC nodes, we can replace the $\regC$ code bounds by the actual code rate in \eqref{rate2}. In Fig. \ref{rate_threshold_261_code} we plot the $\regC$ coding rate (using markers), and the SC upper bound and and the achievable bound of the corresponding family of codes for $(2,6)$ and $(2,7)$ base DDs. Results for $(2,8)$ and $(2,15)$ base DDs can be found in Fig. \ref{rate_threshold_281_4}. Observe that, with the proposed component codes, we are able to perform at least as good as the achievable bound of the corresponding family of block component codes. In some cases, e.g. the $(2,8)$ base DD, the achievable bound is significantly outperformed. Recall that  for the $(2,8)$ base DD the rate bounds in Fig. \ref{rate_threshold_higher}(a) are loose. While for the $(2,7)$ and $(2,15)$ codes the SC bound is attained except for large values of $\fraction$, for the $(2,6)$ and $(2,8)$ ensembles results suggest that there is still room for improving the component code design.

Finally, in the same figures, we highlight those points for which, asymptotically, the $\regC$ ensemble with the proposed linear component codes under P-PD operates closer to channel capacity than the base LDPC code ensemble under PD. For both the $(2,6)$, $(2,7)$, and the $(2,8)$ base DDs we were able to find such points. For the $(2,15)$ ensemble, the minimum gap to capacity obtained is slightly above  the one of the base LDPC code ensemble under PD (0.0743 and 0.0623 respectively). 

\subsection{Growth Rate of the Weight Distribution}

\begin{figure}[h!]
\begin{tabular}{cc}
\hspace{0cm}\includegraphics{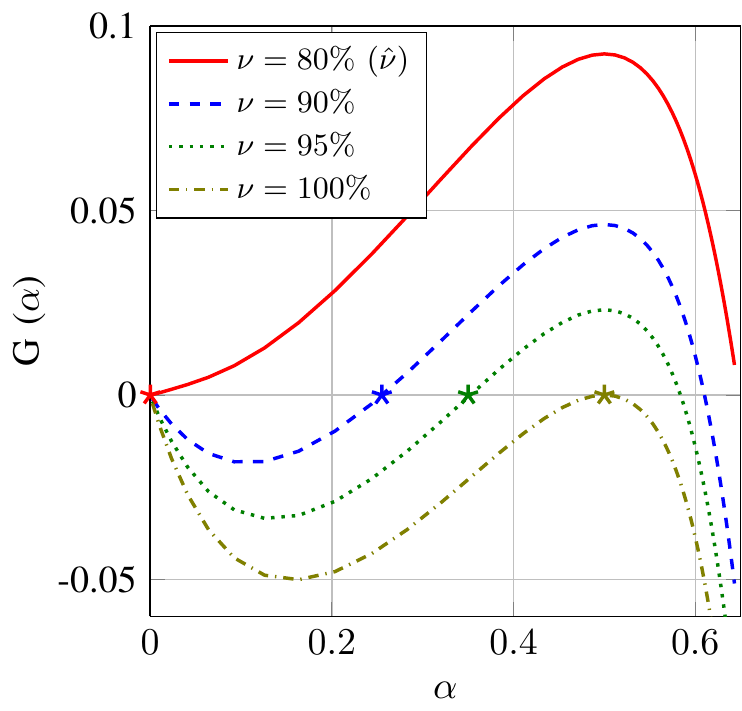} & \includegraphics{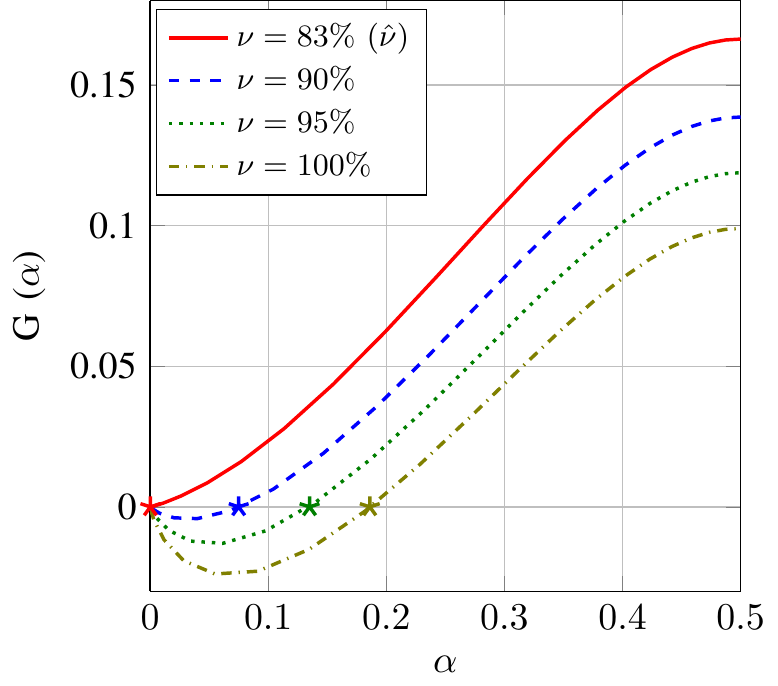}\\
(a) & (b)
\end{tabular}
\caption{In Fig. \ref{weight_distribution} (a), we plot the weight spectral shape $G(\alpha)$ in \eqref{weight} of the $\regC$ ensemble for a $(2,6)$ base DD and with Code R-I as component code. In Fig. \ref{weight_distribution} (b), we plot the same quantity for the $\regC$ ensemble for a $(2,7)$ base DD and with Code R-II as component code (b).}
\label{weight_distribution}
\end{figure}

Upon selecting a specific block code, we can compute the weight spectral shape $G(\alpha)$ in \eqref{weight} using the tools proposed in \cite{Paolini13}. In Fig. \ref{weight_distribution}, we plot $G(\alpha)$ for different values of $\fraction$, computed for the $(2,6)$ base DD with Code R-I as component code (Fig. \ref{weight_distribution} (a)) and the $(2,7)$-regular base DD with Code R-III as component code (Fig. \ref{weight_distribution} (b)). Recall that the critical exponent codeword weight ratio is defined as $\hat{\alpha}\triangleq \inf\{\alpha\geq0 | G(\alpha)\geq 0\}$. In the plots, we highlight $\hat{\alpha}$ with a star. By Lemma \ref{lemma3}, we have $\hat{\alpha}=0$ at $\fraction=\hat{\fraction}$. As $\fraction$ grows, $\hat{\alpha}$ grows, too, and it achieves its maximum at $\fraction=1$. These results indicate that there is a trade-off between the gap to capacity and $\hat{\alpha}(\fraction)$, the critical exponent codeword weight ratio.  As an example, we include values of both quantities in Table \ref{alphatradeoff}  for the  $(2,6)$-regular base DD with Code R-I as component code.

\begin{table}[h!]
\centering
\caption{$\hat{\alpha}, ~\epsilon^*$ and Gap to capacity for different values of $\fraction$, computed for the $(2,6)$-base DD with Code R-I component codes }\label{alphatradeoff}
\begin{tabular}{c|c|c|c}
$\fraction$ & $\alpha$  & $\th$ & Gap to capacity  \\
\hline
$80\%$    & 0           &  0.768  & 0.0987 \\
$87.5\%$ & 0.2049 &  0.788  & 0.1287 \\
$90\%$    & 0.2556 &  0.792  & 0.1413 \\   
$92.5\%$ & 0.3038 &  0.797  & 0.1530 \\
$95\%$    & 0.3526 &  0.801  & 0.1657 \\   
$97.5\%$ & 0.4056 &  0.806  & 0.1773 \\
$100\%$  & 0.6078 &  0.809  & 0.1910 \\  
\end{tabular}
\end{table}

\subsection{Extension to irregular GDLPC code ensembles}\label{irr}

\begin{figure}[t!]
\center
\includegraphics{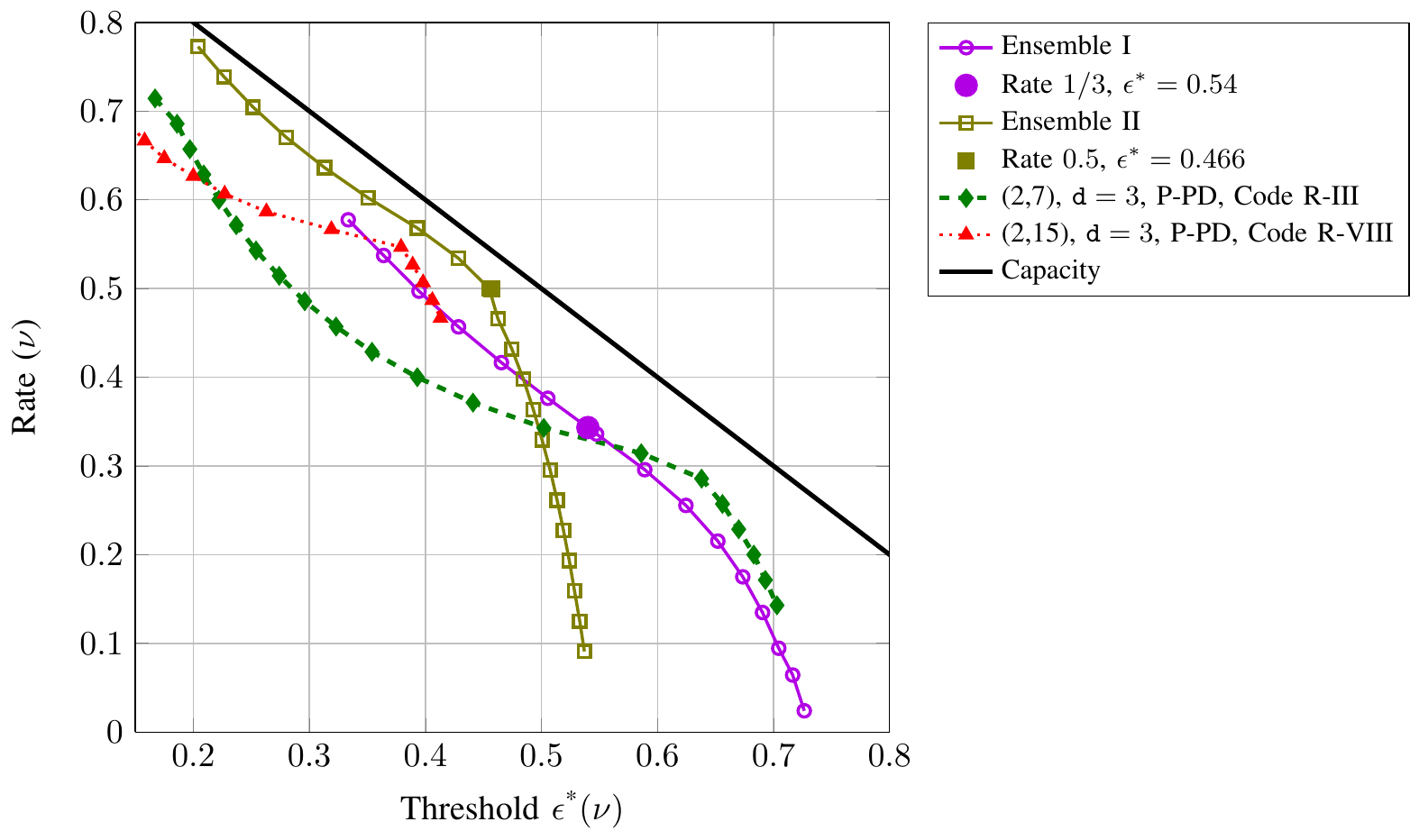}
\caption{P-PD asymptotic threshold and coding rates for different regular and irregular GLDPC code ensembles with varying fraction $\fraction$ of GC nodes in the graph.}\label{Irre}
\end{figure}

To finish this section, we present some further examples using GLDPC code ensembles with irregular DD. Note that the initial conditions in \eqref{diff_gldpc_l_ini}-\eqref{diff_gldpc_l_ini4} of the P-PD asymptotic analysis presented in Section \ref{section_four} already consider an arbitrarily irregular DD, and hence the methodology presented is directly applicable to irregular GLDPC code ensembles. As an example, here we discuss two irregular GLDPC code ensembles:
\begin{itemize}
\item \emph{Ensemble I} \cite{Paolini08}. Rate $1/3$, $\lambda(x) = 0.2 x + 0.7118 x^2 + 0.0882 x^4$, $\fraction^*=0.6719$ and Hamming $(7,4)$ component codes. Using ML decoding at GC nodes, the reported threshold is 0.540.
 \item \emph{Ensemble II} \cite{Guan17}. Rate $1/2$, $\lambda(x) = 0.80 x^2 + 0.01 x^5+ 0.01 x^7 + 0.18 x^{9}$, $\fraction^*=0.40$ and Hamming $(15,11)$ component codes. Using ML decoding at GC nodes, the reported threshold is 0.466. 
\end{itemize}
These ensembles have been constructed using numerical-constrained optimization methods. In Fig. \ref{Irre} we show the results of the P-PD asymptotic analysis when we vary $\fraction$ around the fraction $\fraction^*$ defined above for each case. Observe first that in both cases our results are consistent with the thresholds computed in \cite{Paolini08,Guan17}. In addition, they show that the gap to capacity for Ensemble II can be reduced if we slightly reduce the ensemble rate, i.e. by reducing $\fraction$ to roughly $35$\% instead of $40\%$. For Ensemble I,  the gap to capacity is indeed minimized at exactly the point predicted in \cite{Paolini08}. For comparison, we have included $(2,X)$-regular GLDPC code ensembles with the same check node degrees (and thus same graph density) as Ensembles I and II. Observe that while Ensemble II significantly outperforms the rate-threshold tradeoff of the $(2,15)$-GLDPC code ensemble with Code R-VIII as component code, the $(2,7)$-regular GLDPC code with Code R-III as component code approximately attains threshold $0.540$ at rate $\rate=1/3$, but can reduce the gap to capacity as we decrease the coding rate.



\section{Random puncturing}\label{rateadaptation}

We have proposed the P-PD algorithm as a flexible model to analyze beyond-BD decoding algorithm at GC nodes. Observe that for the P-PD algorithm, the evaluation of the coding rate and the iterative decoding threshold are decoupled problems. This provides a flexible analysis framework that allows the exploration of additional techniques to modify the designs presented above and further reduce the gap to capacity. In this section and the following one, we consider two relevant examples. Specifically, in this section we consider the use of random puncturing to accommodate the coding rate by dropping the transmission of a fraction of coded bits \cite{Mitchell15}. In the next section, a simple model of doubly-generalized LDPC (DG-LDPC) code ensembles is analyzed \cite{Wang06,Yige09, Paolini10}.



\begin{figure}[h!]
\begin{tabular}{cc}
\hspace{0cm}\includegraphics{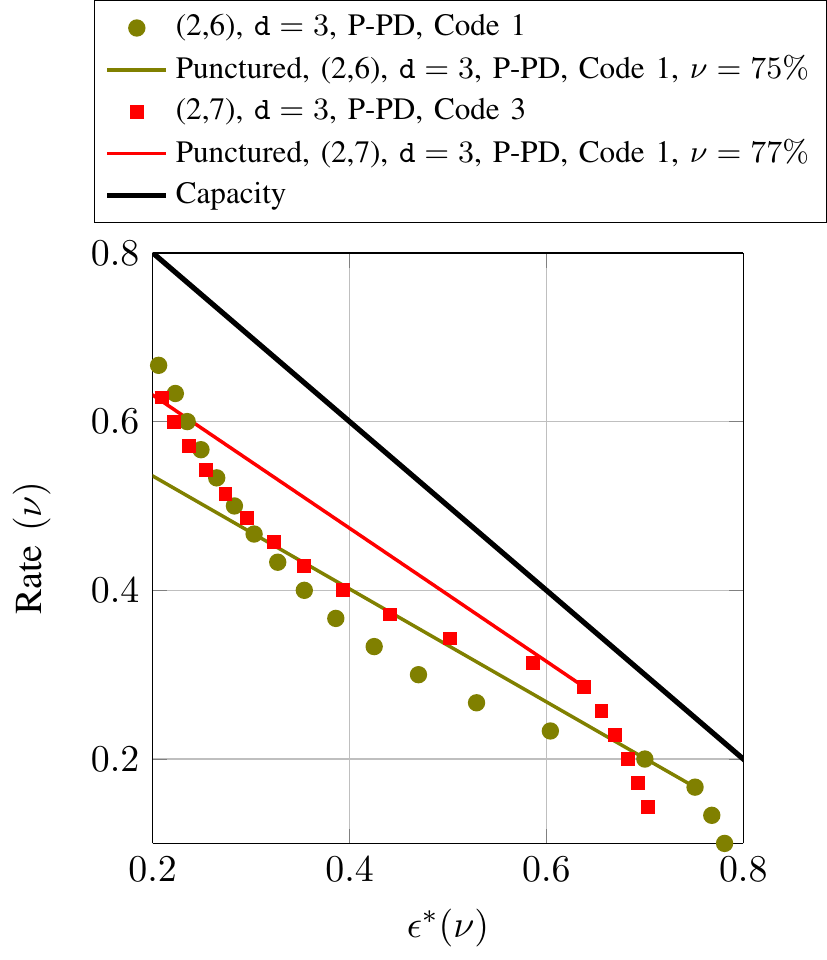} &\hspace{-1cm} \includegraphics{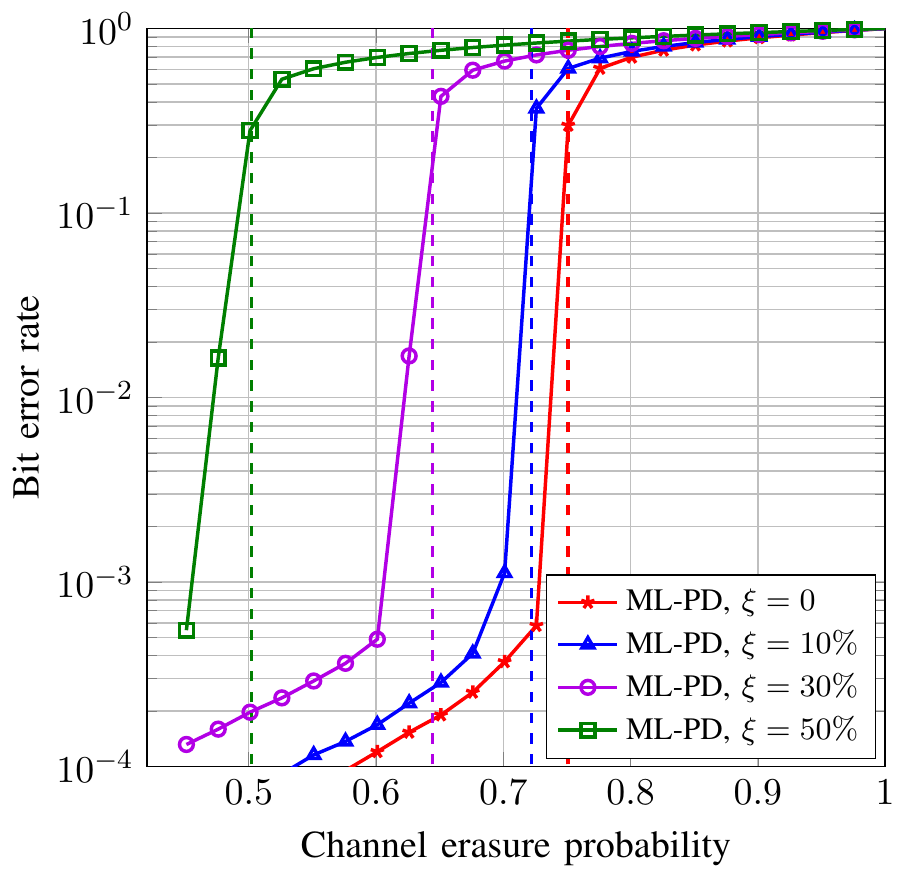} \\
(a) & (b)
\end{tabular} 
\caption{ In Fig. \ref{puncturing} (a), we show with markers the $\regC$ threshold-rate curve for the $(2,6)$ and  the $(2,7)$ base DDs and component codes R-1 and R-3, respectively. Solid lines indicate the evolution of the rates and thresholds of the punctured ensemble for a fixed $\fraction$ as we increase the puncturing fraction $\xi$. In Fig. \ref{puncturing} (b), we show the simulated P-PD performance for the $(2,6)$ base DD with component codes  R-1, $\n=10000$ bits, and different values of the puncturing rate $\xi$. Vertical dashed lines indicate the thresholds predicted by \eqref{th_puncturing}. }\label{puncturing}
\end{figure}

As illustrated in \cite{Mitchell15}, a linear code is \emph{punctured} by removing a set of columns from its generator matrix. After puncturing at random a fraction $\xi$ of the coded bits in the $\regC$  ensemble, the resulting coding rate is 
\begin{align}\label{rate_puncturing}
\rate(\fraction,\xi) = \frac{\rateC}{1-\xi}, ~~ \xi \in[0, 1),
\end{align}
where we recall that $\rateC$ denotes the coding rate of the original $\regC$ ensemble. In \cite{Mitchell15}, the authors derive a simple analytic expression for the iterative belief propagation (BP) decoding threshold of a randomly punctured LDPC code ensemble on the binary erasure channel (BEC). Following their proof, it can be verified that the same results apply to a randomly punctured GLDPC code ensemble. The result reads as follows. Given a $\regC$ ensemble with iterative decoding threshold $\epsilon^*(\fraction)$, the threshold $\epsilon^*(\fraction,\xi)$ of the GLDPC ensemble that follows by randomly puncturing a fraction $\xi$ of the coded bits is related to the unpunctured case as follows:
\begin{align}\label{th_puncturing}
\th(\fraction,\xi) = 1 - \frac{1-\epsilon^*(\fraction)}{1-\xi}.
\end{align}
Observe that the larger the unpunctured threshold $\epsilon^*(\fraction)$ is, the larger the threshold of the punctured ensemble will be. In this regard, we can think of the design of a punctured GLDPC ensemble as a two stage process: First, the GLDPC code ensemble can be designed by choosing $\fraction$ to minimize the gap to capacity. Second, for a fixed $\fraction$, we can analyze the overall gap to capacity as we increasing the code rate by combining \eqref{rate_puncturing} and \eqref{th_puncturing}. We perform this experiment  in Fig. \ref{puncturing} (a) for the $(2,6)$ and the $(2,7)$ base DDs and component codes R-I and R-III, respectively. With markers we show the $\regC$ threshold-rate curve as we increase the fraction of GC nodes in the graph. Solid lines indicate the evolution of the rate and threshold of the punctured ensemble for fixed $\fraction$ as we increase the puncturing fraction $\xi$. Observe that with puncturing it is possible to increase the coding rate and obtain an iterative decoding threshold that is closer to capacity than those obtained by the original $\regC$ ensemble. The accuracy of the predicted threshold can be observed in Fig. \ref{puncturing} (b), where  we include both the threshold predicted by \eqref{th_puncturing} (dashed lines) and  the simulated P-PD performance for the $(2,6)$ base DD with component code R-I, $\n=10000$ bits, and  different values of the puncturing rate $\xi$ (solid lines). We note that, once we introduce puncturing, the SC upper bound in \eqref{scgldpc} is not applicable anymore.

%
%
%

\section{Doubly-generalized LDPC codes}\label{doubly}

\begin{figure}[t]
\centering
\begin{tabular}{cc}
\includegraphics{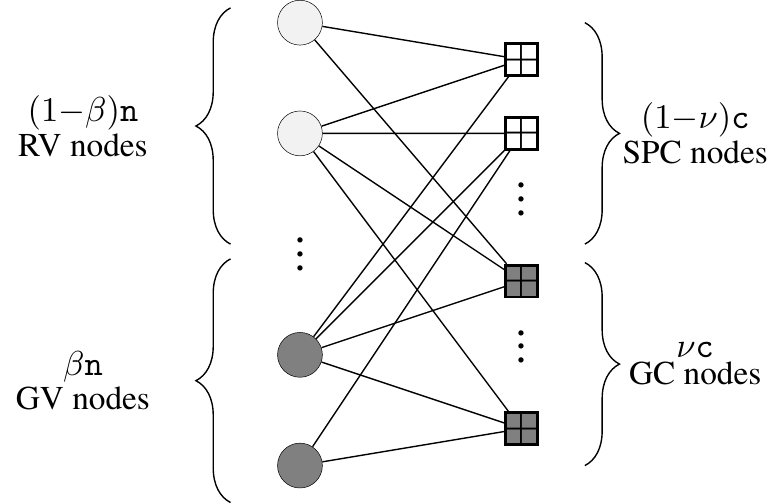}
\end{tabular}
\caption{Tanner graph of the DG-LDPC code ensemble.} \label{DGLDPCgraph}
\end{figure}

A different technique that can potentially help to find a better balance between coding rate and threshold is the inclusion of generalized variable nodes, giving rise to a doubly-generalized LDPC code ensemble \cite{Wang06}. In this section we develop an example with a simple class of a DG-LDPC ensemble. We modify the $\regC$ ensemble by replacing a certain fraction $\beta$ of regular variable  (RV) nodes by generalized variable (GV) nodes, see Fig. \ref{DGLDPCgraph}. Degree-$J$ RV nodes in the $\regC$ graph can be seen as rate $1/J$ repetition code of block length $J$, where the input to the repetition code represents one bit of the DG-LDPC codeword.  On the other hand, degree-$J$ GV nodes are characterized by a $(J,\rowsv)$ linear block code, where the input to the variable component code represents $\rowsv$ bits of the DG-LDPC codeword. Thus, the total block length of the DG-LDPC code ensemble is $\n'=(1-\beta) \n + \beta \n \rowsv$, where $\n$ is the number of variable nodes (both RV and GV) in the graph. In the following, we will assume $J=3, \rowsv=2$ and the following generator matrix for GV nodes:
\begin{align}\label{ggv}
\text{G} = 
\begin{pmatrix}
1 &1 &0 \\
0 &1 &1
\end{pmatrix}.
\end{align}

Thus, each GV node encodes two bits of the DG-LDPC codeword. Denote this ensemble by $\mathcal{C}_{3,K,\fraction,\beta}$. If the component codes at GC nodes are linear block codes with a $\rows$-row parity check matrix, an easy calculation shows that the coding rate of the ensemble is
\begin{align}\label{ratedc}
\rateDC= 1 - (1-\rate_0) \left( \frac{1+(\rows-1)\fraction}{1+ \beta} \right).
\end{align}
As before, we characterize the component codes at GC nodes by the triple $(\dist,p_{\dist},p_{\dist+1})$. Furthermore, the code associated with the generator matrix \eqref{ggv} has minimum distance $2$ and can only decode erasure patterns of weight one. 

\subsection{Decoding via P-PD} \label{DG-LDPC-PPD}

Suppose we use a random sample of the $\mathcal{C}_{3,K,\fraction,\beta}$ code ensemble to transmit over a BEC($\epsilon$). RV nodes are removed from the graph with probability $1-\epsilon$. Regarding GV nodes, we have to consider the following three scenarios:
\begin{itemize}
\item With probability $(1-\epsilon)^2$ the two DG-LDPC coded bits are correctly received and the GV node can be removed from the graph.
\item With probability $2\epsilon(1-\epsilon)$, only one of the two coded bits is received. Since the node is only encoding one unknown bit, note that we can replace the GV node in the graph by a degree-2 RV node. 
\item With probability $\epsilon^2$ the GV node remains in the graph as a degree-$3$ GV node. 
\end{itemize}
Decoding will be performed via P-PD. Since the code spanned by \eqref{ggv} can only decode one error, during the P-PD procedure every GV node needs to lose at least two edges before it can be removed from the graph. Further, once it loses one edge, it can be replaced by a degree-$2$ RV node. Hence, a small modification is required at step 2) in the P-PD Algorithm in Section \ref{Peeling}. Now, it reads as follows: 

\begin{itemize}
\item[] 2) Remove from the Tanner graph the check node with the index drawn in Step 1). Further remove all connected RV nodes, connected degree-$2$ GV nodes and all attached edges. \end{itemize}

\subsection{Degree Distribution and Asymptotic Analysis}\label{DDDG}
While no change is needed to describe the evolution of the check nodes of the residual DG-LDPC code ensemble during P-PD, additional definitions at the variable side are needed to tackle both RV nodes and GV nodes. Let $L_{r2}^{(\ell)}$ and $L_{r3}^{(\ell)}$ represent the total number of edges in the graph connected to RV nodes of degree $2$ and $3$, respectively, after iteration $\ell$ of the decoder. Further let $L_{g3}^{(\ell)}$ be the total number of edges in the graph connected to GV nodes of degree $3$. 
\theorem{\label{ThP-PDDG}
Consider a BEC with erasure probability $\epsilon$ and assume we use elements of the  $\mathcal{C}_{3,K,\fraction,\beta}$ code ensemble for transmission. If we use P-PD  with parameters $(\dist,p_{\dist},p_{\dist+1})$, then the DD of the residual graph at iteration $\ell$ converges  in the sense of \eqref{convergence} to
\begin{align}
L_{r2}^{(\ell)}/\edges&\rightarrow l_{r2}^{(\tau)},\\
L_{r3}^{(\ell)}/\edges&\rightarrow l_{r3}^{(\tau)},\\
L_{g3}^{(\ell)}/\edges&\rightarrow l_{g3}^{(\tau)},\\
R_{pj}^{(\ell)}/\edges&\rightarrow  r_{pj}^{(\tau)}, ~~ j\in\{1,\ldots,K\}\\
R_{cj}^{(\ell)}/\edges&\rightarrow r_{cj}^{(\tau)}, ~~ j \in\{1, \dots, K\}\text{ and } j\notin\{\dist,\dist+1\}\\
\hat{R}_{cj}^{(\ell)}/\edges&\rightarrow \hat{r}_{cj}^{(\tau)}, ~~j\in\{\dist,\dist+1\} \\
\bar{R}_{cj}^{(\ell)}/\edges&\rightarrow \bar{r}_{cj}^{(\tau)} , ~~j\in\{\dist,\dist+1\}
\end{align}
where $l_{r2}^{(\tau)}$, $l_{r3}^{(\tau)}$, $l_{g3}^{(\tau)}$ $r_{pj}^{(\tau)}$, $r_{cj}^{(\tau)}$, $\hat{r}_{cj}^{(\tau)}$, $\bar{r}_{cj}^{(\tau)}$, $\tau=\frac{\ell}{\edges}\in[0,\sum_{i=1}^{J}l^{(\tau)}_i/i]$ are the solutions to the system of differential equations given by \eqref{diff_gldpc_l1}-\eqref{diff_gldpc_rcbar} using $a^{(\tau)} = (3 l_{r3}^{(\tau)} + 2 l_{r2}^{(\tau)} + l_{g3}^{(\tau)} ) / e\t$ and 
\begin{align}\label{dgldpc_diff1} 
\frac{\text{d}l_{r2}^{(\tau)}}{\text{d}\tau}&=2\left( \frac{ l^{(\tau)} _{g3}-l^{(\tau)} _{r2} }{e^{(\tau)}}\right) \left(P_{p1}^{(\tau)} + \sum_{w = 1}^{\dist+1} w P_{c w}^{(\tau)}  \right)\\ \label{dgldpc_diff2}
\frac{\text{d}l_{r3}^{(\tau)}}{\text{d}\tau}&=-\frac{3 l^{(\tau)} _{r3} }{e^{(\tau)}}\left(P_{p1}^{(\tau)} + \sum_{w = 1}^{\dist+1} w P_{c w}^{(\tau)}  \right)\\\label{dgldpc_diff3}
\frac{\text{d}l_{g3}^{(\tau)}}{\text{d}\tau}&=-\frac{3  l^{(\tau)} _{g3} }{e^{(\tau)}}  \left(P_{p1}^{(\tau)} + \sum_{w = 1}^{\dist+1} w P_{c w}^{(\tau)}  \right),
\end{align}
Here, $e^{(\tau)}$, $P_{p1}^{(\tau)}$ and $P_{c w}^{(\tau)}$ are defined in \eqref{defe}, \eqref{defp1}, and \eqref{defpj}, respectively. The initial conditions of the system of differential equations in \eqref{diff_gldpc_l1}-\eqref{diff_gldpc_rcbar} and \eqref{dgldpc_diff1}-\eqref{dgldpc_diff3} are given by
\begin{align} \label{ini_dg_r2}
l_{g3}^{(0)} &= \epsilon^{2} \beta,\\ \label{ini_dg_l}
l_{r3}^{(0)} &= \epsilon (1-\beta), \\\label{ini_dg_r1}
l_{r2}^{(0)} &= 4\beta\epsilon(1-\epsilon)/3 
\end{align}
and by \eqref{diff_gldpc_ini_rp}-\eqref{diff_gldpc_l_ini4} evaluated at $\epsilon' =  \epsilon (1+\beta(1-\epsilon)/3)$.
}
 \begin{IEEEproof}
See Appendix \ref{app2}.
\end{IEEEproof}

\subsection{Results for the $(3,6)$ and $(3,7)$ base DDs}
\begin{figure}
\center
\includegraphics{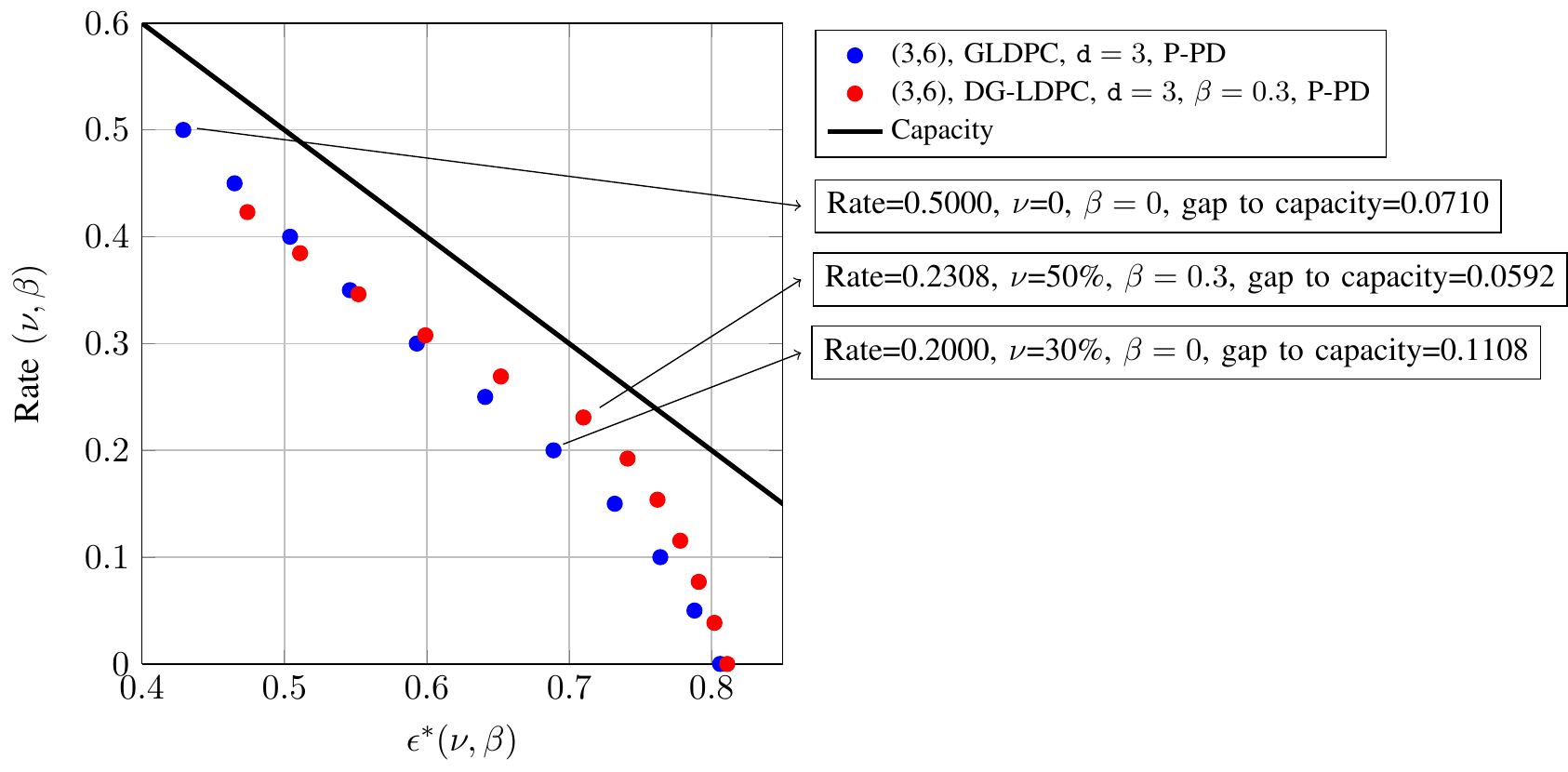}
\caption{$\mathcal{C}_{3,K,\fraction,\beta}$ coding rate for a  (3,6) base DD with Code R-I as component code, GV nodes constructed using the generator matrix in  \eqref{ggv}, and $\beta = 0.3$.}\label{dgldpc36}
\end{figure}

\begin{figure}
\center
\includegraphics{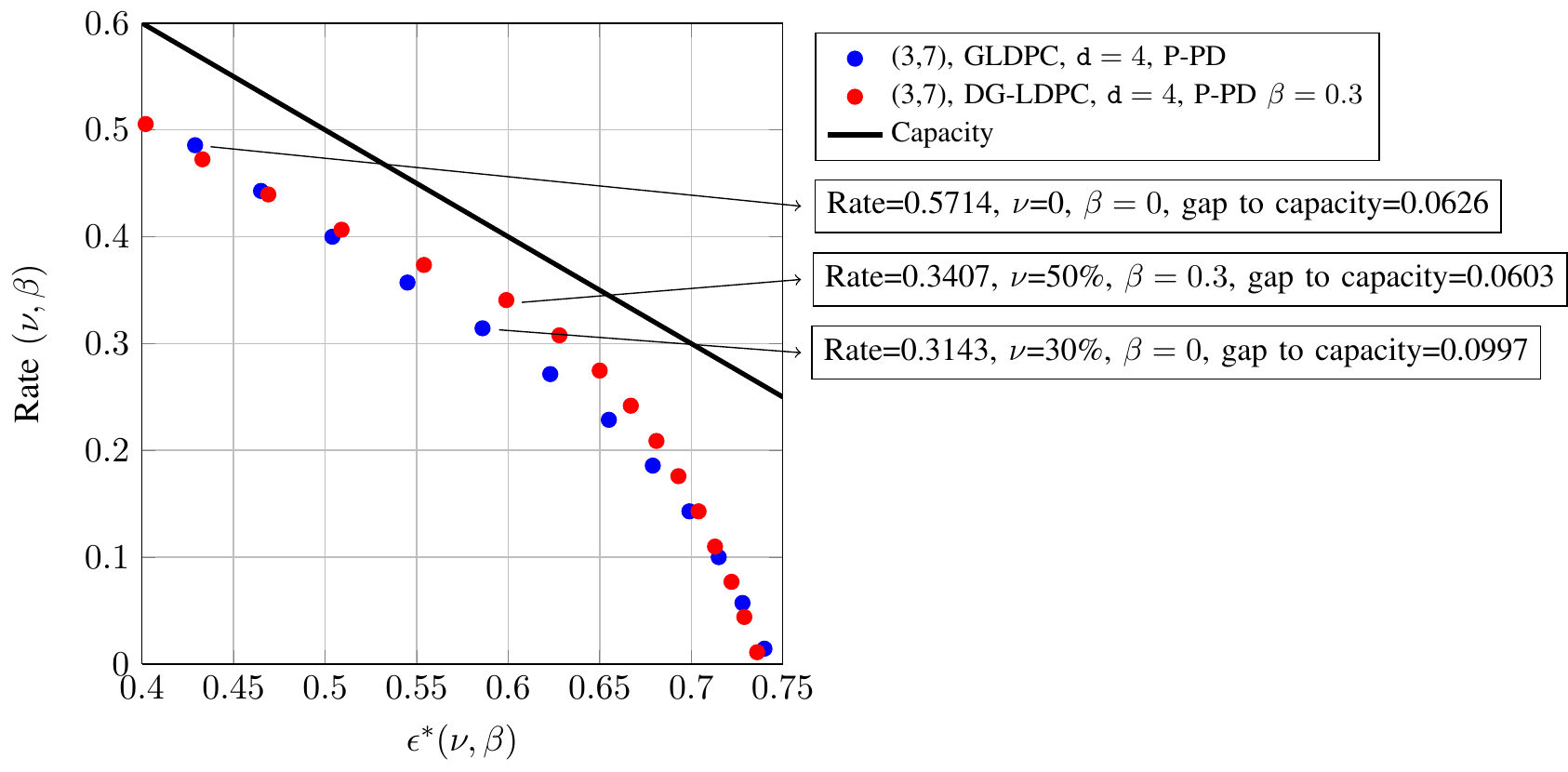}
\caption{$\mathcal{C}_{3,K,\fraction,\beta}$ coding rate for (3,7) base DD, Code R-III component code, GV nodes constructed using the generator matrix in  \eqref{ggv}, and $\beta = 0.3$.}\label{dgldpc37}
\end{figure}

Fig. \ref{dgldpc36} shows the computed rate-threshold curve parametrized by $\fraction$  for both the $\mathcal{C}_{3,K,\fraction,\beta}$ ensembles, both with $\beta=0$, i.e., when the code graph has no generalized variable nodes, and with $\beta=0.3$. We use a $(3,6)$ base DD with code R-I (see Table \ref{tablecodes}) as component code. While in the former case the minimun gap to capacity is achieved for the base LDPC code ensemble (with a gap to capacity of 0.0710), by using a certain amount of generalized variable nodes we are able to reduce this gap to 0.0592. Further, since all variable nodes in the graph have degree $3$, by Lemma \ref{lemma3}, for any value of $\fraction$ the code ensemble has a minimum distance that grows linearly with the block length. Fig. \ref{dgldpc37} shows similar results for a $(3,7)$ base DD with Code R-III as component code. 

%

\section{Conclusions and Future Work}\label{end}

We proposed the P-PD algorithm as a flexible and efficient decoding algorithm that allows us to easily incorporate  ML-decoded GC nodes with specific properties into the asymptotic analysis and still maintain a random definition of the graph degree distribution. Using P-PD, asymptotic analysis of the GLDPC ensemble is carried out by a simple generalization of the original PD analysis by Luby et al. in \cite{Luby01}. The only information required about the component code and its decoding method is the fraction of decodable erasure patterns of a certain weight. We consider a class of GLDPC code ensembles characterized by a regular base DD where we include a certain fraction of GC nodes, and we study the tradeoff between iterative decoding threshold, coding rate and minimum distance. We have shown that  one can find a fraction of GC nodes required that reduces the original gap to capacity and yields  a GLDPC ensemble with linear growth of the minimum distance w.r.t. the block length. Finally, we show how the P-PD analysis can be combined with additional techniques to find a better balance between coding rate and asymptotic gap to capacity. In particular, we consider random puncturing and the use of generalized variable nodes. We would like to emphasize that, in the proposed analysis framework, the evaluation of both coding rate and of iterative decoding threshold are decoupled problems. Consequently, broader classes of component codes or improved decoding methods at GC nodes can be incorporated in a systematic way.

Future lines of work include the analysis of GLDPC codes with regular base DD and a certain fraction of GC nodes in the finite-length regime. Due to their regularity of the DD, we expect such codes to possess a robust finite-length behavior compared to GLDPC code designs proposed in the literature, characterized by capacity-achieving DDs.

\appendices
\section{Wormald's Theorem and the proof of Theorem \ref{ThP-PD}}\label{app0}

Proving Theorem \ref{ThP-PD} is tantamount to showing that the conditions of Wormald’s theorem are satisfied \cite{Wormald}. In this case, Theorem \ref{ThP-PD} follows directly from \eqref{martingale_diff} and \eqref{martingale_p} below.

\subsection{Wormald's theorem \cite{Wormald}}\label{app0wormald}

Let $\{Z^{(\ell)}(a)\}_{a\geq 1}$ be a $d$-dimensional discrete-time Markov random process with state space $\{0,1,\ldots,\lfloor a\alpha\rfloor\}^d$ for $\alpha>0$ and $\ell\in\mathbb{N}_{+}$ denotes the time index. Further let $Z^{(\ell)}_{i}(a)$, $i =1,\dots,d$ denote the $i$-th component of $Z^{(\ell)}(a)$. Let $\mathcal{D}$ be some open connected bounded set containing the closure of
\begin{align}\label{Wormald_1}
\left\{(z_{1},...,z_{d}):~ P\left(\frac{Z^{(0)}_{i}(a)}{a} = z_{i}, 1\leq i \leq d\right) > 0 \text{ for some $a$}\right\}.
\end{align}
We define the stopping time $\ell_{D}$ to be the smallest time index $\ell$ such that  
\begin{align}
( Z^{(\ell_{D})}_{1}(a)/a, ..., Z^{(\ell_{D})}_{d}(a)/a ) \notin \mathcal{D}
\end{align}
Furthermore, let $f_i(\cdot)$, $i=1,\ldots,d$, be functions from $\mathbb{R}^{d+1}$ to $\mathbb{R}$. Assume that the following conditions are satisfied:
\begin{enumerate}
\item (Boundedness) There exists a constant $\nu$ such that for all $ i = 1, \dots d$, $ \ell =0,\dots, \ell_{\mathcal{D}}-1$ and $a \geq 1$,
\begin{align*}
\left| Z^{(\ell+1)}_{i}(a) - Z^{(\ell)}_{i}(a) \right| \leq \nu.  
\end{align*}
\item (Trend functions) For all $i =1, \dots, d$, $ \ell =0,\dots, \ell_{\mathcal{D}}-1$ and $a \geq 1$,
\begin{align*}
\mathbb{E}  \left[ Z^{(\ell+1)}_{i}(a)/a - Z^{(\ell)}_{i}(a)/a \Big| Z^{(\ell)}(a)/a \right]  = f_{i} \left(\ell/a, Z^{(\ell)}_{1}(a)/a, ..., Z^{(\ell)}_{d}(a)/a \right)+\mathcal{O}(1/a).
\end{align*}
\item (Lipschitz continuity) Each function $f_{i}(\cdotp )$, $i =1, \dots, d$,  is Lipschitz continuous on $\mathcal{D}$. Namely, for any pair $b,c \in \mathcal{D}$ that belongs to such intersection, there exists a constant $\kappa $ such that
\begin{align*}
| f_{i}(b) - f_{i}(c)| \leq \kappa \sum\limits_{j = 1}^{d+1} |b_{j} - c_{j}|.
\end{align*}
\end{enumerate}

Under these conditions, the following holds:
\begin{itemize}
\item The system of differential equations 
\begin{align}\label{martingale_diff}
\frac{\partial z_{i}}{\partial \tau} = f_{i}(\tau, z_{1},...,z_{d}), ~~i = 1,...,d,
\end{align} 
has a unique solution for any initial condition  $(b_{1},...,b_{d}) \in \mathcal{D}$.
\item There exists a strictly positive constant $\zeta$ such that
\begin{align}\label{martingale_p} 
P\left(\left|{Z^{(\ell)}_{i}(a)}/a - z_{i}(\ell/a) \right| > \zeta a^{-\frac{1}{6}} \right) = \mathcal{O}\left(\text{e}^{-\sqrt{a}}\right) 
\end{align} 
for $i = 1,...,d$ and $0 \leq t \leq t_{\mathcal{D}}$, where $z_{i}(\ell/a)$ is the solution to \eqref{martingale_diff} for 
\begin{align}\label{martingale_z}
b_{i} = \mathbb{E}[ Z_{i}^{(0)} (a)]/a, \quad i=1,\ldots,d.
\end{align}
\end{itemize}
The result in \eqref{martingale_p} states that any realization of the process $Z_{i}^{(t)}(a)$ concentrates around the solution predicted by \eqref{martingale_diff} in the limit as $a\rightarrow\infty$. In the next subsection we show that this theorem is suitable to describe the expected GLDPC graph evolution of the P-PD.

\subsection{Expected graph evolution under P-PD}\label{app0graph}

To analyze the asymptotic behavior of the $\regC$ ensemble under P-PD using Wormald's theorem, we identify the Markov random process $Z^{(\ell)}(a)$ in the previous section by the random process $\mathcal{G}^{(\ell)}(\edges)$, where
\begin{align}
\mathcal{G}^{(\ell)}(\edges) =\left\{L_{i}^{(\ell)}, R_{pj}^{(\ell)}, R_{cj}^{(\ell)}, \hat{R}_{c\dist}^{(\ell)},  \bar{R}_{c\dist}^{(\ell)}, \hat{R}_{c(\dist+1)}^{(\ell)},  \bar{R}_{c(\dist+1)}^{(\ell)} \right\}_{\substack{i=1,\ldots,J\\j=1,\dots, \dist-1, \dist+2, \dots, K}}
\end{align}
namely $\mathcal{G}^{(\ell)}(\edges)$ is the random process that contains all terms in the DD of the residual graph after $\ell -1 $ iterations. Note that any  component in $\mathcal{G}^{(\ell)}(\edges)$ belongs to the set $\{0, 1, \ldots, \edges\}$, and recall that $\edges$ is the number of edges in the original GLPDC graph. Thus, $\edges$  will play the role of the parameter $a$. In this subsection we prove that the evolution of $ \mathcal{G}^{(\ell)}(\edges)$ under P-PD satisfies the three conditions of Wormald's theorem stated in the previous subsection. We start by computing the conditional expected evolution of all elements in $\mathcal{G}^{(\ell)}(\edges)$ after one P-PD iteration.    We define the following normalized quantities:
\begin{align} \label{normt}
&\tau \triangleq \frac{\ell}{\edges}, ~~ l^{(\ell)} _i \triangleq \frac{L^{(\ell)} _i}{\edges}, ~~r^{(\ell)} _{pj} \triangleq \frac{R^{(\ell)} _{pj}}{\edges}, ~~r^{(\ell)} _{cj} \triangleq \frac{R^{(\ell)} _{cj}}{\edges}, ~~ \hat{r}^{(\ell)} _{c\nu} \triangleq \frac{\hat{R}^{(\ell)} _{c\nu}}{\edges},  ~~\bar{r}^{(\ell)} _{c\nu} \triangleq \frac{\bar{R}^{(\ell)} _{c\nu}}{\edges}, 
\end{align}
for $i\in\{1,\ldots,J\}, j\in\{1,\dots, \dist-1, \dist+2, \dots, K\}$ and $\nu\in\{\dist,\dist+1\}$. We have that
\begin{align}
r^{(\ell)} _{c\nu}=\hat{r}^{(\ell)} _{c\nu}+\bar{r}^{(\ell)} _{c\nu}, ~~ 	\nu=\dist,\dist+1,\\ 
e^{(\ell)} \triangleq \sum_{i=1}^{J}l^{(\ell)} _i= \sum_{j=1}^{K}[r^{(\ell)} _{pj}+r^{(\ell)} _{cj}] , 
\end{align}
and $e\t$ is the fraction of edges remaining in the residual graph at time $\ell$. The P-PD process starts at $\ell=0$, after BEC transmission and initialization. The following relation holds between the quantities defined above at $\ell=0$ and the $\regC$ DD described in Section \ref{codes}:
\begin{align} \label{diff_gldpc_l_ini1}
\mathbb{E}[l_{i}^{(0)}]  &= \epsilon \lambda_i, \\\label{diff_gldpc_ini_rp1}
\mathbb{E}[r_{pj}^{(0)}]  &= \sum_{\alpha \geq j} \rho_{p\alpha} {\alpha-1 \choose j-1} \epsilon ^{j} (1-\epsilon)^{\alpha -j}, \\\label{diff_gldpc_ini_rc1}
\mathbb{E}[r_{cj}^{(0)}] &= \sum_{\alpha \geq j} \rho_{c\alpha} {\alpha-1 \choose j-1} \epsilon ^{j} (1-\epsilon)^{\alpha -j}, 
\end{align}
for $i=1,\ldots J$ and $j=1,\ldots,K$, where the expectation is computed w.r.t. the $\regC$ ensemble and the channel output. Upon initialization, every degree-$\dist$ GC node is tagged as decodable with probability $p_\dist$, and every degree-$(\dist+1)$ GC node is tagged as decodable with probability $p_{\dist+1}$. Recall that all GC nodes with degree less than $\dist$ are decodable and, by assumption, all GC nodes with degree more than $\dist+1$ are not decodable. We thus have the following initial conditions
\begin{align}\nonumber
\mathbb{E}[\hat{r}_{cj}^{(0)}] &= p_{j}\mathbb{E}[r_{cj}^{(0)}], \\
\mathbb{E}[\bar{r}_{cj}^{(0)}] &= (1-p_j)\mathbb{E}[r_{cj}^{(0)}] , ~~ j = \dist, \dist+1 \label{diff_gldpc_l_ini31}.
\end{align}
 
The equations \eqref{diff_gldpc_l_ini1}-\eqref{diff_gldpc_l_ini31} correspond to the initial conditions in \eqref{martingale_z}. Observe that since the largest GC degree is $K$ and the largest variable node degree is $J$, the graph loses at most $JK$ edges per iteration. This is an upper bound on the absolute variation of any component in $\mathcal{G}^{(\ell)}(\edges)$ between two consecutive iterations. Hence, Condition 1) of Wormald's theorem is satisfied.

Suppose we observe $\mathcal{G}^{(\ell)}(\edges)$. To derive the conditional expectations  in Condition 2) of Wormald's Theorem, the so-called trend functions, we have to average among every possible scenario that we can observe after a P-PD iteration. According to Step 1) in  Algorithm \ref{P-PD}, we chose at random a decodable check node. Let $P_{p1}\l$ be  the probability of selecting a degree-one SPC node, and let $P_{cj}\l$ denote the probability of selecting a decodable degree-$j$ GC node, $j = 1,\dots, \dist+1$. By a simple counting argument, if the check node is selected uniformly at random then 
\begin{align}\label{P_p1}
P_{p1}\l &=\frac{r_{p1}^{(\ell)}}{s\t},\\\label{P_p2}
P_{cj}\l &=\frac{r^{(\ell)} _{cj}/j}{s\t}, ~~~ j< \dist,\\\label{P_p3}
P_{cj}\l &=\frac{ \hat{r}^{(\ell)} _{cj}/j}{s\t}, ~~~ j\in\{\dist,\dist+1\}.
\end{align}
In \eqref{P_p1}-\eqref{P_p3},
\begin{align}\label{P_sum}
s\t &= r_{p1}^{(\ell)} +  \sum_{w =1}^{\dist-1} \frac{r^{(\ell)} _{cw}}{w} + \frac{\hat{r}^{(\ell)} _{c\dist}}{\dist} + \frac{\hat{r}^{(\ell)} _{c(\dist+1)}}{\dist+1} 
\end{align}
is the normalized sum of decodable check nodes at the $\ell$-th iteration. 

\subsubsection{Evolution of left edge degrees in the Tanner graph after one P-PD iteration}
Suppose we observe the residual graph $\mathcal{G}^{(\ell)}$ at iteration $\ell$. Our aim is to evaluate 
\begin{align}\label{expectation_left}
\mathbb{E}\left[L^{(\ell+1)}_i - L^{(\ell) }_i \Big| \mathcal{G}^{(\ell)}(\edges) \right ],
\end{align}
for $i = 1, 2, ..., J$. Given the graph DD $\mathcal{G}^{(\ell)}$, recall that  $P_{p1}\l$ denotes the probability of P-PD selecting a degree-one SPC node in the current iteration, and $P_{cj}\l$ denotes the probability of selecting a degree-$j$ decodable GC node. We can decompose the expectation in \eqref{expectation_left} according to each possible type of check node to be removed, namely,
\begin{align}\nonumber
\mathbb{E} \left[L^{(\ell+1)}_i - L^{(\ell) }_i \Big|  \mathcal{G}^{(\ell)}(\edges)  \right ] &= P_{p1}\l \mathbb{E} \left[L^{(\ell + 1)}_i - L^{(\ell) }_i \Big| \mathcal{G}^{(\ell)}(\edges) , \text{Deg}_{p1} \right ] \\\label{new}
& +  \sum_{w = 1}^{\dist+1 }P_{cw}\l \mathbb{E} \left[L^{(\ell +1 )}_i - L^{(\ell) }_i \Big|  \mathcal{G}^{(\ell)}(\edges),  \text{Deg}_{cw} \right ],
\end{align}
where $\text{Deg}_{p1}$ indicates that the P-PD removes a degree-one SPC node from the graph, and $\text{Deg}_{cw}$ indicates that P-PD removes a degree-$w$ decodable GC node from the graph. Computing the expectation in the first case is similar to the derivation carried out in \cite{Luby01} for PD with LDPC ensembles. Indeed probability that the edge adjacent to the removed degree-one SPC node has left degree $i$ is $l_i\t/e\t$. In such a case, after deleting this variable node, the graph loses $i-1$ additional edges adjacent to this variable node, so
\begin{align} \label{diff_left}
\mathbb{E} \left[L^{(\ell + 1)}_i - L^{(\ell) }_i \Big|  \mathcal{G}^{(\ell)}(\edges), \text{Deg}_{p1} \right ] = - \frac { i l^{(\ell)} _i }{e\l}.
\end{align}
When the P-PD decoder removes a decodable degree-$w$ GC node, this node is connected to  $w$ variable nodes that are also removed from the residual Tanner graph, along with their connected edges (assuming the graph does not have double edges). Note that left degrees of the $w$ edges connected to the removed GC node are, in general, not independent. Let $X_u\in\{1,\ldots,J\}$ the RV that indicates the left degree of the $u$-th edge, $u=1,\ldots,w$. Arbitrarily, we can decompose the joint probability of $X_1,\ldots,X_w$ as follows
\begin{align}\label{expansion}
P(X_1,\ldots,X_w)=P(X_1)P(X_2|X_1)P(X_3|X_1,X_2)\ldots P(X_w|X_1,\ldots,X_{w-1}).
\end{align}
While $P(X_1=x_1)=l_{x_1}\l/e\l$, $x_1=1,\ldots,J$, the conditional distribution of $X_2$ given $X_1$ is given by
\begin{align}
P(X_2=x_2|X_1=x_1)=\left\{\begin{array}{cc}\displaystyle \frac{l_{x_2}\t}{e\l-1/\edges} & x_2\neq x_1\\\\ \displaystyle\frac{l_{x_2}\l-1/\edges}{e\l-1/\edges} & x_2=x_1\end{array}\right.,
\end{align}
for $x_1,x_2\in\{1,\ldots,J\}$, where the $1/\edges$ terms appear due to the fact that the DD has to be reparameterized after we condition on $X_1=x_1$. The above expression can be generalized to any of the factors in \eqref{expansion} as follows:
\begin{align}\label{eq73}
P(X_u=x_u|X_1=x_1,\ldots,X_{u-1}=x_{u-1})&=\frac{\displaystyle l_{x_u}\l-\frac{\sum_{u'=1}^{u}\mathbbm{I}[x_{u'}=x_u]}{\edges}}{\displaystyle e\t-\frac{u-1}{\edges}}\nonumber\\
&=\left(\frac{l_{x_u}\l}{e\l}-\frac{\sum_{u'=1}^{u}\mathbbm{I}[x_{u'}=x_u]}{e\l\edges}\right)\frac{e\l \edges }{e\l \edges -(u-1)}.
\end{align}
Note that $e\t\edges$ is the number of edges in the graph at time $\ell$.  Since $u\leq w< J$ and $J$ is a constant independent of $\edges$, the second factor in \eqref{eq73} is of order $1-\mathcal{O}(1/\edges)$. Thus
\begin{align}
P(X_1=x_1,\ldots,X_w=x_w)=\prod_{u=1}^{w}\left(\frac{l_{x_u}\l}{e\l}-\frac{\sum_{u'=1}^{u}\mathbbm{I}[x_{u'}=x_u]}{e\l\edges}\right)+\mathcal{O}(1/\edges),
\end{align}
using again that  $w\leq\dist+1\leq J$ where $J$ is a constant independent of $\edges$, and that $l_{x_u}\l/e\l$ is independent of $\edges$, we can write \eqref{expansion}  as follows
\begin{align}\label{joint}
P(X_1=x_1,\ldots,X_w=x_w)=\prod_{u=1}^{w}\frac{l_{x_u}\l}{e\l}+\mathcal{O}(1/\edges).
\end{align}
Thus, the joint probability distribution of the left degrees of $w$ edges connected to a degree-$w$ GC node asymptotically factorizes as $\edges\to\infty$ and  the number of edges with left degree-$i$ connected to the removed GC node can be roughly described by a binomial RV with parameter $l\l _i/e\l$. Hence, we obtain
\begin{align}\label{eq76}
\mathbb{E} \left[L^{(\ell +1 )}_i - L^{(\ell) }_i \Big| \mathcal{G}^{(\ell)}(\edges), \text{Deg}_{cw}\right ]=- \frac { i w l^{(\ell)} _i }{e\t}+\mathcal{O}(1/\edges).
\end{align}
Combining \eqref{eq76} and \eqref{diff_left} with \eqref{new}, we obtain
\begin{align}\label{fi}
\mathbb{E} \left[L^{(\ell + 1)}_i - L^{(\ell) }_i \Big| \mathcal{G}^{(\ell)}(\edges) \right ] = - \frac{i l^{(\ell)} _i }{e\t}\left(P_{p1}\l + \sum_{w = 1}^{\dist + 1} w P_{c w}\l  \right)+\mathcal{O}(1/\edges) \triangleq f_i( \mathcal{G}^{(\ell)}(\edges)/\edges)+\mathcal{O}(1/\edges).
\end{align}
Note that $f_i(\mathcal{G}^{(\ell)}(\edges)/\edges)$ depends on every component in $\mathcal{G}^{(\ell)}$, normalized by $\edges$. Observe that $f_i(\mathcal{G}^{(\ell)}(\edges)/\edges)$ in \eqref{fi} is of the form required by Condition 2) of Wormald's theorem. 

\subsubsection{Evolution of right edge degrees in the Tanner graph after one P-PD iteration} Our goal now is to evaluate 
\begin{align}\nonumber
&\mathbb{E}\left[R_{pj}^{(\ell+1)} - R_{pj}^{(\ell)}  \Big| \mathcal{G}^{(\ell)}(\edges) \right], ~~j =1,\ldots,K,  \\\nonumber
&\mathbb{E}\left[R_{cj}^{(\ell+1)} - R_{cj}^{(\ell)}\Big| \mathcal{G}^{(\ell)}(\edges) \right], ~~j=1, \ldots,K \text{ and } j\notin\{\dist,\dist+1\} \\\nonumber
&\mathbb{E}\left[\hat{R}_{cj}^{(\ell+1)} - \hat{R}_{cj}^{(\ell)}\Big| \mathcal{G}^{(\ell)}(\edges) \right], ~~j\in\{\dist,\dist+1\}\\\nonumber
&\mathbb{E}\left[\bar{R}_{cj}^{(\ell+1)} - \bar{R}_{cj}^{(\ell)}\Big| \mathcal{G}^{(\ell)}(\edges) \right],  ~~j\in\{\dist,\dist+1\}\nonumber
\end{align}

As before, we evaluated these terms by conditioning on the type of check node to be removed at the current P-PD iteration. Using \eqref{joint}, the average number of edges removed from the graph after a degree-$w$ GC node is  removed is given by $\Delta_w\t\triangleq wa\l+\mathcal{O}(1/\edges)$, where $a\l = \sum i l_i^{(\ell)} / e\l$. Among those, $w$ are connected to the same degree-$w$ GC node, i.e. they have right degree $w$. Consider the remaining $\Delta_w-w$ edges. Following a similar argument as in \eqref{joint}, it can be shown that the joint probability distribution of their right degree asymptotically factorizes as $\edges\rightarrow\infty$ and that the deviation in the finite case is dominated by $\mathcal{O}(1/\edges)$ terms. By taking $w=1$, the same arguments hold for the case where decoder removes a degree-1 SPC node.  In addition to this results, in order to evaluate the expected variation in the number of edges of certain right degree we also have to take into account that, when we remove one edge from the graph, we modify the right degree of the rest of edges still connected to the same SPC/GC node. For example, if one of the edges that are removed from the graph has right SPC degree $j$, after deleting such edge the graph loses $j$ edges with right SPC degree $j$ and gains $j-1$ edges with right SPC degree $j-1$.  

Following the above arguments,  conditioned on $\mathcal{G}^{(\ell)}(\edges)$, the expected change in the number of edges with right SPC degree $j$ is given by the following expression
\begin{align} \label{Rchange}
&\mathbb{E} \left[R^{(\ell +1 )}_{pj} - R^{(\ell) }_{pj} \Big|  \mathcal{G}^{(\ell)}(\edges) \right ] \nonumber\\
&= P_{p1}\l \left((r^{(\ell)}_{p(j+1)} - r^{(\ell)} _{pj} ) \frac {  j (a(\ell) -1) }{e\l} -\mathbbm{I}[j=1]\right) +  \sum_{w = 1}^{\dist +1}P_{cw}\l (r^{(\ell)}_{p(j+1)} - r^{(\ell)} _{pj} ) \frac { j  (wa(\ell) -w) }{e\l}+\mathcal{O}(1/\edges)\nonumber\\
&\triangleq g_{pj}(\mathcal{G}^{(\ell)}/\edges) +\mathcal{O}(1/\edges).
\end{align}
It can be further shown that the expected variation in the number of edges of right GC degree $j$ with $j\neq{\dist,\dist+1}$ satisfies
\begin{align}
&\mathbb{E} \left[R^{(\ell + 1 )}_{cj} - R^{(\ell) }_{cj} \Big| \mathcal{G}^{(\ell)}(\edges) \right ] \nonumber\\
&= P_{p1}\l \left((r^{(\ell)}_{c(j+1)} - r^{(\ell)} _{cj} ) \frac {  j (a(\ell) -1) }{e\l} \right)  + \sum_{w = 1}^{\dist +1 }P_{cw}\l \left((r^{(\ell)}_{c(j+1)} - r^{(\ell)} _{cj} ) \frac { j  (wa(\ell) -w) }{e\l}-w\mathbbm{I}[j=w]\right) +\mathcal{O}(1/\edges)\nonumber\\
&\triangleq g_{cj}(\mathcal{G}^{(\ell)}/\edges) +\mathcal{O}(1/\edges).
\end{align}
To analyze the expected change in the number of edges connected to decodable and not decodable GC nodes of degree $\dist$ and $\dist+1$, we have to take into account that if a non-decodable degree-$(d+2)$ GC node loses one edge, it becomes decodable with probability $p_{\dist+1}$. Similarly, if a non-decodable degree-$(d+1)$ GC node loses one edge, it becomes decodable with probability $p_{\dist}$. 
Also note that if a decodable GC node of degree $\dist+1$ loses one edge, it becomes a decodable GC node of degree $\dist$ with probability 1. It follows that the expected change in the fraction of edges connected to decodable and not decodable GC nodes of degree $j = \dist,\dist+1$, are given by 
\begin{align} \nonumber
&\mathbb{E} \left[\hat{R}^{(\ell + 1 )}_{cj} - \hat{R}^{(\ell) }_{cj} \Big| \mathcal{G}^{(\ell)}(\edges) \right ] \\\nonumber
&= P_{p1}\l \left((p_{j} \bar{r}^{(\ell)}_{c(j+1)}+\hat{r}^{(\ell)}_{c(j+1)} - \hat{r}^{(\ell)} _{cj} ) \frac {  j (a(\ell) -1) }{e\l} \right)\\\nonumber
&+ \sum_{w = 1}^{j +1}P_{cw}\l \left((p_{j} \bar{r}^{(\ell)}_{c(j+1)}+\hat{r}^{(\ell)}_{c(j+1)}  - \hat{r}^{(\ell)} _{cj} ) \frac { j (wa(\ell) -w) }{e\l}-w \mathbbm{I}[w=j]\right) +\mathcal{O}(1/\edges)\\\label{fistright}
&\triangleq \hat{g}_{cj}(\mathcal{G}^{(\ell)}/\edges) +\mathcal{O}(1/\edges)
\end{align}
\begin{align} \nonumber
&\mathbb{E} \left[\bar{R}^{(\ell + 1 )}_{cj} - \bar{R}^{(\ell) }_{cj} \Big| \mathcal{G}^{(\ell)}(\edges) \right ] \\\nonumber
&= P_{p1}\l \left(((1-p_{j}) \bar{r}^{(\ell)}_{c(j+1)} - \bar{r}^{(\ell)} _{cj} ) \frac {  j (a(\ell) -1) }{e\l} \right) \\\nonumber
&+\sum_{w = 1}^{j +1}P_{cw}\l \left(((1-p_{j}) \bar{r}^{(\ell)}_{c(j+1)}  - \bar{r}^{(\ell)} _{cj} ) \frac { j  (wa(\ell) -w) }{e\l} -w \mathbbm{I}[w=j] \right) +\mathcal{O}(1/\edges)\nonumber\\\label{lastexp}
&\triangleq \bar{g}_{cj}(\mathcal{G}^{(\ell)}/\edges) +\mathcal{O}(1/\edges)
\end{align}
Note that $\bar{R}^{(\ell)}_{c(\dist+2)} = R^{(\ell)}_{c(\dist+2)}$ and $\hat{R}^{(\ell)}_{c(\dist+2)} = 0$.  Further, observe that \eqref{fi}-\eqref{lastexp} are of the form required by Condition 2) of Wormald's theorem.  

\subsubsection{On the Lipschitz continuity of the trend functions in \eqref{fi}-\eqref{lastexp}}

Condition 3) of  Wormald's theorem requires that the trend functions in \eqref{fi}-\eqref{lastexp}  are Lipschitz in the set of all possible DDs.  First, we note that if we would  restrict the  P-PD to remove only decodable check nodes (either degree-1 SPC nodes or GC nodes of one particular degree), then \eqref{fi}-\eqref{lastexp}  are still valid  by simply setting the corresponding probabilities $P_{p1}\l$ and $P_{cj}\l$, $j=1,\ldots,\dist+1$ to either zero or one. In such a case, \eqref{fi}-\eqref{lastexp}  are equal up to a multiplicative constant to the PD trend functions for LDPC codes in \cite{Luby01},  hence they are Lipschitz continuous. When we drop the restriction to remove one particular type of decodable check node, then the trend functions in \eqref{fi}-\eqref{lastexp}  are convex the combinations of Lipschitz continuous functions, with the coefficients given by the functions $P_{p1}\l$ and $P_{cj}\l$, $j=1,\ldots,\dist+1$ in \eqref{P_p1}-\eqref{P_p3}, which are also Lipschitz continuous (note their similarity in form with \eqref{diff_left}, which is Lipschitz continuous \cite{Luby01}). Since they are all bounded functions, we conclude that Condition 3) of  Wormald's theorem is also satisfied.

 
\section{Proof of Theorem \ref{ThP-PDDG}}\label{app2}


The proof of Theorem \ref{ThP-PDDG} closely follows that  of Theorem \ref{ThP-PD} given in Appendix \ref{app0}. As before, it is sufficient to show that the conditions of Wormald’s theorem are satisfied. 
Following the definitions  given in Section \ref{DDDG}, the left DD of the residual graph of the $\mathcal{C}_{3,K,\fraction,\beta}$ code ensemble during P-PD has three components: the number of edges connected to degree-$2$ or degree-$3$ RV nodes ($L_{r2}^{(\ell)}$ and $L_{r3}^{(\ell)}$ respectively), and the number of edges connected to degree-$3$ GV nodes ($L_{g3}^{(\ell)}$). The right DD of the residual graph has the same elements as those defined for the $\regC$ ensemble in Appendix \ref{app0graph}. Thus, the DD of the residual graph is defined by the random process
\begin{align}
\mathcal{G}^{(\ell)}(\edges) =\left\{L_{r2}^{(\ell)},L_{r3}^{(\ell)}, L_{3g}^{(\ell)},  R_{pj}^{(\ell)}, R_{cj}^{(\ell)}, \hat{R}_{c\dist}^{(\ell)},  \bar{R}_{c\dist}^{(\ell)}, \hat{R}_{c(\dist+1)}^{(\ell)},  \bar{R}_{c(\dist+1)}^{(\ell)} \right\}_{\substack{j=1,\dots, \dist-1, \dist+2, \dots, K}}.
\end{align}
We define 
\begin{align}\label{newdd}
l_{r2}^{(\ell)}\triangleq\frac{L_{r2}^{(\ell)}}{\edges}, ~ l_{r3}^{(\ell)}\triangleq\frac{L_{r3}^{(\ell)}}{\edges}, ~ l_{g3}^{(\ell)}\triangleq\frac{L_{g3}^{(\ell)}}{\edges}.
\end{align}
After P-PD initialization, i.e. $\ell=0$, it can be shown that
\begin{align} \label{ini_dg_r22}
\mathbb{E} \left[l_{g3}^{(0)}\right] &= \epsilon^{2} \beta,\\ \label{ini_dg_l2}
\mathbb{E} \left[l_{r3}^{(0)}\right] &= \epsilon (1-\beta), \\\label{ini_dg_r12}
\mathbb{E} \left[l_{r2}^{(0)}\right] &= 4\beta\epsilon(1-\epsilon)/3.
\end{align}
To evaluate \eqref{ini_dg_r12}, we compute the average number of GV nodes for which one of the two DG-LDPC coded bits is received. According to the generator matrix in \eqref{ggv}, GV nodes can be viewed as degree-$2$ variable nodes.  Based on \eqref{ini_dg_r22}-\eqref{ini_dg_r12}, the average fraction of edges remaining in the graph after P-PD initialization is
\begin{align}\label{ini_dg_r3}
\epsilon' = \epsilon (1-\beta) + 4\beta\epsilon(1-\epsilon)/3 +  \epsilon^{2} \beta= \epsilon \left(1+\frac{\beta(1-\epsilon)}{3}\right).
\end{align}
We can further determine expected initial conditions of the right DD of the residual graph after P-PD initialization by using \eqref{diff_gldpc_ini_rp} and \eqref{diff_gldpc_l_ini3} and replacing $\epsilon$ by $\epsilon'$. 

By following a similar procedure as in Appendix \ref{app0graph}, it can be shown that conditioned, on $\mathcal{G}^{(\ell)}(\edges)$, the expected variation in $L_{r2}^{(\ell)},L_{r3}^{(\ell)}$, and $L_{3g}^{(\ell)}$ after one P-PD iteration is given by 
\begin{align}\label{dgldpc_diff1} 
\mathbb{E} \left[L^{(\ell + 1)}_{r3} - L^{(\ell) }_{r3} \Big| \mathcal{G}^{(\ell)} \right ] &= -\frac{3 l^{(\ell)} _{r3} }{e^{(\ell)}}\left(P_{p1}\l + \sum_{w = 1}^{\dist+1} w P_{c w}\l  \right)+\mathcal{O}(1/\edges), \\\label{dgldpc_diff2}
\mathbb{E} \left[L^{(\ell + 1)}_{r2} - L^{(\ell) }_{r2} \Big| \mathcal{G}^{(\ell)} \right ] &=  \left( \frac{2 l^{(\ell)} _{g3} }{e^{(\ell)}} - \frac{2  l^{(\ell)} _{r2} }{e^{(\ell)}}   \right) \left(P_{p1}\l + \sum_{w = 1}^{\dist+1} w P_{c w}\l  \right) +\mathcal{O}(1/\edges),  \\\label{dgldpc_diff3}
\mathbb{E} \left[L^{(\ell + 1)}_{g3} - L^{(\ell) }_{g3} \Big| \mathcal{G}^{(\ell)} \right ] &= -\frac{3  l^{(\ell)} _{g3} }{e^{(\ell)}}  \left(P_{p1}\l + \sum_{w = 1}^{\dist+1} w P_{c w}\l  \right) +\mathcal{O}(1/\edges),
\end{align}
where $e(\ell)=l_{r3}^{(\ell)}+l_{g3}^{(\ell)}+l_{g3}^{(\ell)}$ and $P_{p1}\l$ and $P_{c w}\l$ are given in \eqref{defp1} and \eqref{defpj} respectively. In \eqref{dgldpc_diff2}, we have used that that if a degree-$3$ GV node loses one edge, then the graph loses $3$ edges with left GV degree $3$ and gains $2$ edges with left RV degree $2$. The conditional expected variation of the right DD of the residual graph can be computed  using  \eqref{Rchange}-\eqref{lastexp} by taking $a^{(\ell)} = (3 l_{r3}^{(\ell)} + 2 l_{r2}^{(\ell)} + l_{g3}^{(\ell)} ) / e^{(\ell)}$. Finally, proving that the conditions in Wormald's Theorem hold follows by the same arguments as in  the proof of Theorem \ref{ThP-PD} in Appendix \ref{app0}. 

\section{Generator matrices of reference Codes}\label{app1}

Reference codes have been found by performing an exhaustive search over the database \cite{Grassl:codetables,Grassl06}, which implements MAGMA \cite{Magma} to design block codes with the largest minimum distance. 

\vspace{0.5cm}
\emph{Code R-I}: Rate-$1/2$ Hamming $(6,3)$ linear block code with generator matrix

\begin{align}
\text{G}_{\text{R-I}} = 
\begin{pmatrix}
1 &0 &0 &1 &1 &0 \\
0 &1 &0 &1 &0 &1 \\
0 &0 &1 &0 &1 &1
\end{pmatrix}
\end{align}

\vspace{0.5cm}
\emph{Code R-II}: Rate-$1/3$  Cordaro-Wagner $2$-dimensional repetition code of length $6$ with generator matrix

\begin{align}
\text{G}_{\text{R-II}} = 
\begin{pmatrix}
1 &1 &1 &1 &0 &0 \\
0 &0 &1 &1 &1 &1 
\end{pmatrix}
\end{align}

\vspace{0.5cm}
\emph{Code R-III:} Rate-$4/7$ Hamming (7,4) code with generator matrix

\begin{align}
\text{G}_{\text{R-III}} = 
\begin{pmatrix}
1 &1 &1 &0 &0 &0 &0 \\
1 &0 &0 &1 &1 &0 &0 \\
0 &1 &0 &1 &0 &1 &0 \\
1 &1 &0 &1 &0 &0 &1
\end{pmatrix}
\end{align}

\vspace{0.5cm}
\emph{Code R-IV:} Rate-$3/7$ linear block code with generator matrix

\begin{align}
\text{G}_{\text{R-IV}} = 
\begin{pmatrix}
0 &1 &1 &1 &1 &0 &0 \\
1 &1 &0 &1 &0 &1 &0 \\
1 &0 &1 &1 &0 &0 &1
\end{pmatrix}
\end{align}


\vspace{0.5cm}     
\emph{Code R-V:} Rate-$1/2$ extended $(7,4)$-Hamming code with extra parity bit, i.e., $(8,4)$ Hamming code. Another example is a Quasi-Cyclic $(8,4,4)$ code with generator matrix 

\begin{align}
\text{G}_{\text{R-V}} = 
\begin{pmatrix}
1 &0 &0 &1 &0 &1 &0 &1 \\
0 &1 &1 &0 &0 &1 &0 &1 \\
0 &1 &0 &1 &1 &0 &0 &1 \\
0 &1 &0 &1 &0 &1 &1 &0
\end{pmatrix}
\end{align}

\vspace{0.5cm}
\emph{Code R-VI:} Rate-$3/8$ cyclic linear block code with generator matrix
 
\begin{align}
\text{G}_{\text{R-VI}} = 
\begin{pmatrix}
1 &0 &0 &1 &1 &0 &0 &1 \\
0 &1 &0 &1 &0 &1 &0 &1 \\
0 &0 &1 &1 &0 &0 &1 &1 
\end{pmatrix}
\end{align}

\vspace{0.5cm}


\emph{Code R-VII:} Rate-$1/4$ Cordaro-Wagner $2$-dimensional repetition code of length 8 with generator matrix

\begin{align}
\text{G}_{\text{R-VII}}=
\begin{pmatrix}
1 &0 &1 &1 &0 &1 &1 &1 \\
0 &1 &0 &0 &1 &1 &1 &1 \\
\end{pmatrix}
\end{align}

\vspace{0.5cm}

\emph{Code R-VIII:} Rate-$11/15$ linear block code with generator matrix 

\begin{align}
\text{G}_{\text{R-VIII}}=
\begin{pmatrix}
0 &1 &0 &1 &0 &1 &1 &0 &0 &0 &0 &0 &0 &0 &0 \\  
0 &0 &0 &1 &0 &1 &0 &0 &1 &0 &0 &0 &0 &0 &0 \\ 
0 &0 &0 &1 &0 &1 &0 &0 &0 &1 &0 &0 &0 &0 &1 \\ 
0 &0 &0 &1 &0 &0 &1 &0 &0 &0 &1 &0 &0 &0 &0 \\ 
0 &0 &0 &1 &0 &0 &1 &0 &0 &0 &0 &1 &0 &0 &1 \\ 
0 &0 &0 &0 &0 &1 &1 &0 &0 &0 &0 &0 &1 &0 &0 \\ 
0 &0 &0 &0 &0 &1 &1 &0 &0 &0 &0 &0 &0 &1 &1 \\ 
0 &0 &0 &0 &0 &0 &1 &1 &0 &0 &0 &0 &0 &0 &1 \\ 
0 &0 &0 &0 &1 &1 &0 &0 &0 &0 &0 &0 &0 &0 &1 \\ 
0 &0 &1 &1 &0 &0 &0 &0 &0 &0 &0 &0 &0 &0 &1 \\ 
1 &0 &0 &1 &0 &1 &1 &0 &0 &0 &0 &0 &0 &0 &1
\end{pmatrix}
\end{align}

\vspace{0.5cm}


\emph{Code R-IX:} Rate-$2/3$ linear block code with generator matrix

\begin{align}
\text{G}_{\text{R-IX}}= 
\begin{pmatrix}
0 &1 &1 &0 &0 &1 &0 &1 &0 &0 &0 &0 &0 &0 &0\\ 
0 &0 &1 &0 &0 &1 &0 &0 &1 &0 &0 &0 &0 &0 &1\\
0 &0 &1 &0 &0 &0 &0 &1 &0 &1 &0 &0 &0 &1 &0\\
0 &0 &1 &0 &0 &1 &0 &0 &0 &0 &1 &0 &0 &1 &0\\
0 &0 &1 &0 &0 &0 &0 &1 &0 &0 &0 &1 &0 &0 &1\\
0 &0 &0 &0 &0 &1 &0 &1 &0 &0 &0 &0 &1 &0 &1\\
0 &0 &0 &0 &0 &1 &1 &0 &0 &0 &0 &0 &0 &1 &1\\
0 &0 &0 &0 &1 &0 &0 &1 &0 &0 &0 &0 &0 &1 &1\\
0 &0 &1 &1 &0 &1 &0 &1 &0 &0 &0 &0 &0 &1 &1\\
1 &0 &1 &0 &0 &0 &0 &0 &0 &0 &0 &0 &0 &1 &1
\end{pmatrix}
\end{align}

 \bibliographystyle{IEEEtran}
\bibliography{allbib}

\begin{thebibliography}{10}
\providecommand{\url}[1]{#1}
\csname url@samestyle\endcsname
\providecommand{\newblock}{\relax}
\providecommand{\bibinfo}[2]{#2}
\providecommand{\BIBentrySTDinterwordspacing}{\spaceskip=0pt\relax}
\providecommand{\BIBentryALTinterwordstretchfactor}{4}
\providecommand{\BIBentryALTinterwordspacing}{\spaceskip=\fontdimen2\font plus
\BIBentryALTinterwordstretchfactor\fontdimen3\font minus
  \fontdimen4\font\relax}
\providecommand{\BIBforeignlanguage}[2]{{%
\expandafter\ifx\csname l@#1\endcsname\relax
\typeout{** WARNING: IEEEtran.bst: No hyphenation pattern has been}%
\typeout{** loaded for the language `#1'. Using the pattern for}%
\typeout{** the default language instead.}%
\else
\language=\csname l@#1\endcsname
\fi
#2}}
\providecommand{\BIBdecl}{\relax}
\BIBdecl

\bibitem{Tanner81}
R.~Tanner, ``A recursive approach to low complexity codes,'' \emph{IEEE
  Transactions on Information Theory}, vol.~27, no.~5, pp. 533 -- 547, Sept.
  1981.

\bibitem{Lentmaier99}
M.~Lentmaier and K.~Zigangirov, ``{On generalized low-density parity-check
  codes based on Hamming component codes},'' \emph{IEEE Communications
  Letters}, vol.~3, no.~8, pp. 248--250, Aug 1999.

\bibitem{Yue07}
G.~Yue, L.~Ping, and X.~Wang, ``{Generalized Low-Density Parity-Check Codes
  Based on Hadamard Constraints},'' \emph{IEEE Transactions on Information
  Theory,}, vol.~53, no.~3, pp. 1058--1079, March 2007.

\bibitem{Liva08}
G.~Liva, W.~Ryan, and M.~Chiani, ``{Quasi-cyclic generalized LDPC codes with
  low error floors},'' \emph{IEEE Transactions on Communications}, vol.~56,
  no.~1, pp. 49--57, January 2008.

\bibitem{Paolini10}
{Paolini, E. and Fossorier, M.P.C. and Chiani, M.}, ``{Generalized and Doubly
  Generalized LDPC Codes With Random Component Codes for the Binary Erasure
  Channel},'' \emph{IEEE Transactions on Information Theory}, vol.~56, no.~4,
  pp. 1651--1672, April 2010.

\bibitem{Mulholland15}
I.~P. Mulholland, E.~Paolini, and M.~F. Flanagan, ``Design of ldpc code
  ensembles with fast convergence properties,'' in \emph{IEEE International
  Black Sea Conference on Communications and Networking, Constanta, Romania},
  May 2015, pp. 53--57.

\bibitem{Mitchell13GLDPC}
D.~Mitchell, M.~Lentmaier, and D.~Costello, ``{On the minimum distance of
  generalized spatially coupled LDPC codes},'' in \emph{Proc. IEEE
  International Symposium on Information Theory (ISIT), Istanbul, Turkey}, July
  2013, pp. 1874--1878.

\bibitem{adr11}
S.~{Abu-Surra}, D.~Divsalar, and W.~E. Ryan, ``Enumerators for protograph-based
  ensembles of {LDPC} and generalized {LDPC} codes,'' \emph{IEEE Transactions
  on Information Theory}, vol.~57, no.~2, pp. 858--886, Feb. 2011.

\bibitem{Lentmaier10GLDPC}
M.~Lentmaier and G.~Fettweis, ``{On the thresholds of generalized LDPC
  convolutional codes based on protographs},'' in \emph{Proc. IEEE
  International Symposium on Information Theory Proceedings (ISIT), Austin,
  USA.}, June 2010, pp. 709--713.

\bibitem{Jian2012}
Y.~Y. Jian, H.~D. Pfister, and K.~R. Narayanan, ``Approaching capacity at high
  rates with iterative hard-decision decoding,'' in \emph{2012 IEEE
  International Symposium on Information Theory Proceedings}, July 2012, pp.
  2696--2700.

\bibitem{Paolini13}
M.~F. Flanagan, E.~Paolini, M.~Chiani, and M.~P.~C. Fossorier, ``Spectral shape
  of doubly-generalized ldpc codes: Efficient and exact evaluation,''
  \emph{IEEE Transactions on Information Theory}, vol.~59, no.~11, pp.
  7212--7228, Nov 2013.

\bibitem{Luby01}
M.~Luby, M.~Mitzenmacher, M.~Shokrollahi, and D.~Spielman, ``Efficient erasure
  correcting codes,'' \emph{IEEE Transactions on Information Theory}, vol.~47,
  no.~2, pp. 569 --584, Feb. 2001.

\bibitem{Urbanke08}
C.~Measson, A.~Montanari, and R.~Urbanke, ``Maxwell construction: The hidden
  bridge between iterative and maximum a posteriori decoding,'' \emph{IEEE
  Transactions on Information Theory}, vol.~54, no.~12, pp. 5277 --5307, Dec.
  2008.

\bibitem{Olmos15}
P.~Olmos, D.~Mitchell, and J.~Costello, D.J., ``{Analyzing the finite-length
  performance of generalized LDPC codes},'' in \emph{2015 IEEE International
  Symposium on Information Theory (ISIT), Hong Kong, China}, June 2015, pp.
  2683--2687.

\bibitem{Burshtein04}
D.~Burshtein and G.~Miller, ``Efficient maximum-likelihood decoding of {LDPC}
  codes over the binary erasure channel,'' \emph{IEEE Transactions on
  Information Theory}, vol.~50, no.~11, pp. 2837 -- 2844, Nov. 2004.

\bibitem{Urbanke08-2}
T.~J. Richardson and R.~Urbanke, \emph{Modern Coding Theory}.\hskip 1em plus
  0.5em minus 0.4em\relax Cambridge University Press, Mar. 2008.

\bibitem{MacKay03}
\BIBentryALTinterwordspacing
D.~J.~C. MacKay, \emph{Information Theory, Inference, and Learning
  Algorithms}.\hskip 1em plus 0.5em minus 0.4em\relax Cambridge University
  Press, 2003. [Online]. Available: \url{http://www.cambridge.org/0521642981}
\BIBentrySTDinterwordspacing

\bibitem{MacWilliams77}
F.~J. MacWilliams and N.~J.~A. Sloane, \emph{The theory of error correcting
  codes}.\hskip 1em plus 0.5em minus 0.4em\relax North-Holland Pub. Co. New
  York, 1977.

\bibitem{Huffman03}
W.~C. Huffman and V.~Pless, \emph{Fundamentals of error-correcting
  codes}.\hskip 1em plus 0.5em minus 0.4em\relax Cambridge, U.K., New York:
  Cambridge University Press, 2003.

\bibitem{Paolini08}
E.~Paolini, M.~Fossorier, and M.~Chiani, ``{On the design of irregular GLDPC
  codes with low error floor over the BEC},'' in \emph{2008 International
  Symposium on Information Theory and Its Applications}, Dec 2008, pp. 1--6.

\bibitem{Guan17}
R.~Guan and L.~Zhang, ``Hybrid hamming gldpc codes over the binary erasure
  channel,'' in \emph{2017 11th IEEE International Conference on
  Anti-counterfeiting, Security, and Identification (ASID)}, Oct 2017, pp.
  130--133.

\bibitem{Mitchell15}
D.~Mitchell, M.~Lentmaier, A.~Pusane, and D.~Costello, ``{Randomly Punctured
  LDPC Codes},'' \emph{IEEE Journal on Selected Areas in Communications},
  vol.~34, no.~2, pp. 408--421, Feb 2016.

\bibitem{Wang06}
Y.~Wang and M.~Fossorier, ``{Doubly Generalized LDPC Codes},'' in \emph{2006
  IEEE International Symposium on Information Theory}, July 2006, pp. 669--673.

\bibitem{Yige09}
------, ``{Doubly Generalized LDPC Codes over the AWGN Channel},'' \emph{IEEE
  Transactions on Communications}, vol.~57, no.~5, pp. 1312--1319, May 2009.

\bibitem{Gallager63}
R.~G. Gallager, \emph{Low Density Parity Check Codes}.\hskip 1em plus 0.5em
  minus 0.4em\relax MIT Press, 1963.

\bibitem{Miladinovic08}
N.~Miladinovic and M.~Fossorier, ``{Generalized LDPC codes and generalized
  stopping sets},'' \emph{IEEE Transactions on Communications}, vol.~56, no.~2,
  pp. 201--212, February 2008.

\bibitem{Wormald}
N.~C. Wormald, ``Differential equations for random processes and random
  graphs,'' \emph{Annals of Applied Probability}, vol.~5, no.~4, pp.
  1217--1235, 1995.

\bibitem{Rich-Sho-Urb-LDPC01}
T.~Richardson, A.~Shokrollahi, and R.~Urbanke, ``Design of capacity-approaching
  irregular low-density parity-check codes,'' \emph{IEEE Transactions on
  Information Theory}, vol.~47, no.~2, pp. 619 --637, Feb. 2001.

\bibitem{Paolini07}
E.~Paolini, M.~Fossorier, and M.~Chiani, ``{Generalized Stability Condition for
  Generalized and Doubly-Generalized LDPC Codes},'' in \emph{2007 IEEE
  International Symposium on Information Theory}, June 2007, pp. 1536--1540.

\bibitem{Grassl:codetables}
M.~Grassl, ``{Bounds on the minimum distance of linear codes and quantum
  codes},'' Online available at \url{http://www.codetables.de}, 2007, accessed
  on 2017-01-07.

\bibitem{Grassl06}
------, ``{Searching for linear codes with large minimum distance},'' in
  \emph{Discovering Mathematics with Magma --- Reducing the Abstract to the
  Concrete}, ser. Algorithms and Computation in Mathematics, W.~Bosma and
  J.~Cannon, Eds.\hskip 1em plus 0.5em minus 0.4em\relax Heidelberg: Springer,
  2006, vol.~19, pp. 287--313.

\bibitem{Magma}
W.~Bosma, J.~Cannon, and C.~Playoust, ``{The Magma Algebra System I: The User
  Language},'' \emph{Journal of Symbolic Computation}, vol.~24, no. 3-4, pp.
  235--265, Oct. 1997.

\end{thebibliography}

\end{document}